\documentstyle[12pt,amstex,amsfonts]{article}

\makeatletter\@addtoreset{equation}{section}\makeatother

\newtheorem{Th}{Theorem}[section]
\newtheorem{Rk}{Remark}[section]
\newtheorem{Lm}{Lemma}[section]
\newtheorem{Co}{Corollary}[section]

\newtheorem{Pn}{Proposition}[section]
\newcommand{\Tr}{\operatorname{Tr}}

\def\1{{1\!\!1}}

\def\R{{I\!\! R}}
\def\N{{I\!\! N}}
\def\Z{{Z\!\!\! Z}}
\def\pl{\partial}

\def\ss{\subset}
\def\sm{\setminus}

\def\1{{1\!\!1}}
\def\a{\alpha}

\def\Tr{\mbox{Tr}}

\def\r{\rho}
\def\b{\beta}
\def\d{\delta}
\def\g{\gamma}

\def\l{\lambda}
\def\om{\omega}
\def\s{\sigma}

\def\p{\phi}
\def\ve{\varepsilon}

\def\D{\Delta}

\def\L{\Lambda}
\def\Om{\Omega}
\def\S{\Sigma}

\def\A{{\cal A}}
\def\AC{{\frak A}}

\def\HC{{\cal H}}

\def\KC{{\cal K}}

\def\LC{{\cal L}}

\def\i{\infty}

\def\be{\begin{equation}}
\def\ee{\end{equation} }

\allowdisplaybreaks

\begin{document}

\font\fifteen=cmbx10 at 15pt \font\twelve=cmbx10 at 12pt

\begin{titlepage}

\begin{center}

\renewcommand{\thefootnote}{\fnsymbol{footnote}}

{\fifteen GIBBS STATE UNIQUENESS FOR ANHARMONIC QUANTUM CRYSTAL
WITH A NONPOLYNOMIAL  DOUBLE-WELL POTENTIAL}

\vspace{1.5 cm}

\setcounter{footnote}{0}
\renewcommand{\thefootnote}{\arabic{footnote}}

\bf{ Alexei L. REBENKO \footnote{On leave of absence from
Institute of Mathematics, Kiev, Ukraine} and Valentin A.
ZAGREBNOV}\footnote{Universit\'e de la M\'editerran\'ee
(Aix-Marseille II)}

\vspace{0.5 cm}

{\twelve Centre de Physique Th{\'e}orique\footnote{Unit{\'e}
Propre de Recherche 7061 }, CNRS-Luminy,  Case 907}

{\twelve F-13288 Marseille -- Cedex 9, France}

\vspace{1.0 cm}

{\bf Abstract}

\end{center}
We construct the Gibbs state for $\nu$-dimensional quantum crystal
with site displacements from $\R^d$ , $d\geq 1$, and with a
one-site \textit{non-polynomial} double-well potential, which has
\textit{harmonic} asymptotic growth at infinity. We prove the
uniqueness of the corresponding {\it Euclidean Gibbs measure}
(EGM) in the \textit{light-mass regime} for the crystal particles.
The corresponding state is constructed via a cluster expansion
technique for an arbitrary temperature $T\geq 0$. We show that for
all $T\geq 0$ the Gibbs state (correlation functions) is analytic
with respect to external field conjugated to displacements
provided that the mass of particles $m$ is less than a certain
value $m_* >0$. The high temperature regime is also discussed.

\vspace{2.0 cm}

\\

\renewcommand{\thefootnote}{\fnsymbol{footnote}}

\end{titlepage}

\setcounter{footnote}{0}
\renewcommand{\thefootnote}{\arabic{footnote}}

\newpage

\vskip1cm


\section{Introduction}

It is generally excepted that for  investigation of different
physical phenomena in quantum crystals one can consider an
infinite system of interacting anharmonic  oscillators, which are
situated in the sites of $\nu$-dimensional lattice $\Z^\nu$. The
heuristic Hamiltonian of  such a system (in the case of 2-body
interaction) has the following form:

\begin{equation}
H \; = \; - \frac{\hbar^2}{2m} \sum_{j \in \Z^\nu} \D_j + \sum_{j
\in \Z^\nu} W(q_{j}) + \frac{J}{2} \sum_{i,j\in\Z^\nu,i\ne j}
Q(q_{i} , q_{j})-\sum_{j \in \Z^\nu}h q_j , \label{1.1}
\end{equation}
where $ m $ is the mass of  particles, the operator $\D_j$
corresponds the kinetic energy of the system and, in fact, is
$d$-dimensional Laplace operator in the one-particle Hilbert space
$ L^{2}(\R^{d}, dq) $, where $ dq $ is the Lebesgue measure on $
\R^{d} $,  $q_j\in\R^d$ for $j\in\Z^\nu$ is displacement of a
particle from its position in the site $j\in\Z^\nu$. For general
case $d\leq\nu\geq 1$. But the most of  results, which where
obtained earlier are for $d=1$. The particles are confined near
their sites by potential $W(q_j)$. A \textit{harmonic} one-site
potential $W_{harm}(q_j)=\frac{1}{2}aq_j^2$,  $a>0$ together with
a \textit{harmonic} two-particle interaction $Q(\cdot,\cdot)$
define a well-known harmonic crystal model (1.1). To produce a
model describing a (ferroelectric) structural phase transition one
usually takes for $W(\cdot)$ a \textit{double-well anharmonic}
potential, keeping  $Q(\cdot,\cdot)$ harmonic (see e.g.
[1,19,24]). For example,
\begin{equation*}
W(q) \; = \frac{1}{4} b(q^2)^2 - \frac{1}{2}a q^2, \quad
a,b\in\R^1_+,
\end{equation*}
or semibounded from below polynomials of higher degree with more
than two equal minima. Here $q^2\equiv q\cdot q$ is the scalar
square of  displacement vector $q\in\R^d$.
 The proof of  existence of phase transition in such kind of systems
for  {\it harmonic} interaction $Q(q_{i} , q_{j})$ were obtained
in [12,15,26,34,40,47,53] for the case

\begin{equation}
Q(q_{i} , q_{j}) = Q_{ij} (q_i-q_j)^2, \label{1.2}
\end{equation}
where $\hat{Q}=(Q_{ij})_{i,j\in\Z^\nu}$ is a matrix of
non-negative harmonic-force constants. The term $h\cdot
q_j=\sum_{\a=1}^d h^{\a}q_j^{\a}$ break the symmetry in the
direction of the external field $h$. The non-uniqueness of Gibbs
states for $h=0$ is proved
 when the mass of the particles $m$ is sufficiently
large and the temperature $T$ is sufficiently low, or $\b=(kT)^{-1}$ is
sufficiently large.

From the other side, another interesting phenomenon, the
\textit{suppression} of the long-range order by strong quantum
fluctuations in such systems was experimentally observed (see,
e.g. [55]) and was discussing long time ago from the physical
point of view, see [51], or the books [1,19]. A rigorous study of
this phenomenon was given in the series of papers. For the $d=1$
displacement and all $\nu\geq 1$ the suppression of the order
parameter for all temperatures including $T=0$ (as soon as the
particle mass is smaller than some threshold: $m < m_*$) was
proved in [58]. Later it was shown in [3] for $d=1$ and in [37-39]
for $d\in\N$, that not only the long-range order but also any
critical behaviour of displacements of particles from the
equilibrium positions are suppressed at all temperatures $T>0$ if
the model is ``strongly quantum'', i.e., if the mass $m$ is small
enough. But a more important problem is to prove the
\textit{uniqueness} of the Gibbs state for such systems for
sufficiently small mass of particles. It would imply the absence
of all critical anomalies in the regime of ``strong quantumness''.
The first step in this direction was made in  [44] for $d=1$,
where for small values of $m$ the convergence of corresponding
cluster expansions was proved for all values of temperature $T\geq
0$. See also [8], where the cluster expansion is constructed by
another and simpler technique but for small $m$, which depends on
the positive $T>0$. The result of [44] has been extended to the
multi-dimensional spin in [27]. This in particular implies that
the {\it order parameter} is always zero if $m < m_*$ for any
values of $T\geq 0$ [43]. It gives, also, \textit{uniqueness} of
the EGM (see [2,31,33] and [7] for detail description of the
Euclidean approach) at least for so-called ``compact'' boundary
conditions for displacements. In particular, this is true if
displacements are reduced to a compact (so-called \textit{compact
spins}). This resumes the results of [44] and [43] about the
\textit{suppression} of the displacement phase transition by the
large quantum fluctuations in \textit{small masse} domain $m<
m_*$. For models with the compact spin the uniqueness was proved
in [9]. Then in [4] the uniqueness of EGM was proved for $d=1$ and
for every fixed inverse temperature $\b<\infty$ if the mass is
less than some threshold: $m< m_*(\b)$. Finally, the
\textit{dependence} of the mass threshold on the temperature was
eliminated  in recent papers [5],[6] for the set of {\it tempered}
EGM.
\begin{Rk}\label{s-larger-1}
These results were obtained  for the case of polynomial anharmonic
one-site potential of the form:
\begin{equation}
W(q)=v(q^2)=\sum_{p = 1}^{s} a_{2p}(q^2)^{p}, \; \; a_{2s} > 0, \;
\; s > 1, \; \; s \in \N. \label{1.3}
\end{equation}
with additional assumption on $ W(q) $  that $v(\cdot)$ is {\it
convex}  on $\R_{+}^{1} $. For the proof of these results the fact
of \textit{smallness} of the parameter $m^{(s-1)/(s+1)}$ for small
$m$, i.e. $s>1$, is crucial. Recall that in this case, after an
appropriate unitary transformation of the Hamiltonian, the
parameter $m^{(s-1)/(s+1)}$ stands in front of the harmonic
interaction term $Q$. So the \textit{light-mass regime}
corresponds to the one-site energy domination and the \textit{weak
interaction} of anharmonic oscillators defined by $W$, that
implies the \textit{uniqueness} of the EGM.
\end{Rk}
On the other hand, let us define $s \geq 0$ by
\begin{equation}\label{s-less}
s = \inf_{\alpha \geq 0} \{\,\alpha :\,
\limsup_{q^2\rightarrow\infty}v(q^2)/(q^2)^{\alpha} < \infty \}.
\end{equation}
If $s \leq 1$, then in the \textit{light-mass domain}
(``\textit{strong quantumness}'') the parameter $m^{(s-1)/(s+1)}$
is not small, i.e. the approach developed for $s > 1$ does not
work. So, the rate of the growth of anharmonic potential is an
important parameter distinguishing different regimes, see
discussion in [43] and [58].
\begin{Rk}\label{s-less-1}
If $s < 1$, then in the light-mass limit the harmonic interaction
term dominates the one-site anharmonic potential, i.e. the system
has to behave close to a Debye harmonic crystal (not proved). In
the marginal case $s = 1$ the both harmonic interaction and
harmonic part of  $\,W$  dominate the anharmonicity. From the
physical point of view the large-distant behaviour of the one-site
anharmonic potential is not well-justified. So, a priori all above
scenario are possible.
\end{Rk}
For example, the paper [54] proposed a model with the one-site
non-polynomial anharmonic potential of the form:
\begin{equation}
W(q) \; = \; \frac{1}{2}a q^2 + b e^{- \d q^2 /2},\quad
a,b,\d\in\R_+^1.\label{1.4}
\end{equation}
corresponding to $s = 1$. It is easy to see that for $a<b\d$ the
minimum of the potential (1.5) is not unique and the system may
undergo a displacement phase transition. For the case of so-called
{\it self-consistent phonon approximation} [1], or for the
\textit{displacement quantum spherical model} it was proved in
[48] for (1.3) and in [54,58] for (1.5). In general case this
follows from [15], see also [47]. On the other hand, in [48] and
[54,58] it is shown that there is a \textit{non-zero} threshold
mass $m_*$ such that the phase transition in the
\textit{displacement quantum spherical model} is suppressed by the
``strong quantumness'' as soon as $m < m_*$.

The aim of the present paper is study the case of the
\textit{quadratic growth}, $s = 1$, of the double-well anharmonic
one-site potential $W$ (\ref{1.4}). We prove that (similar to the
case $s > 1$) there is a \textit{non-zero} mass $m_*$ such that in
the \textit{light-mass} domain: $m < m_*$ the critical behaviour
of the quantum anharmonic system (\ref{1.1}) for $h=0$ is
suppressed for arbitrary dimension of particles displacements
$d\geq 1$ and $\nu\geq 3$. More precisely, we prove the
\textit{uniqueness} of the corresponding Gibbs state. Our result
is based on the proof of convergence of the \textit{cluster
expansions} for the quantum Euclidean Gibbs Measure in the
thermodynamic limit for arbitrary temperature $T\geq 0$ and
sufficiently small mass $m < m_*$ of particles (uniformly in
temperature). Notice that since for $s = 1$ the interaction in the
light-mass limit $m < m_*$ is not negligible (Remark
\ref{s-less-1}), our strategy is different to that for $s > 1$, cf
[44]. In that sense our result supplements the recent proof of the
Gibbs state uniqueness for $s > 1$, see [5,6].

Now there are few other remarks in order to outline \textit{our
strategy} of the proof the uniqueness for the case $s = 1$.

The anharmonic perturbation in (1.5) can be re-written in the
following form
\begin{equation}
V(q) \; = \; b e^{- \d q^2 /2} = b\int d\mu(\a)
e^{i\sqrt{\d}\a\cdot q},\label{1.5}
\end{equation}
where $d\mu(\a)=(2\pi)^{-d/2}e^{-\a^2 /2}d\a$ is the Gaussian
measure on $\R^d$. So anharmonic term (1.6) is a  {\it gentle}
perturbation of the quantum harmonic system in the sense of
Albeverio--H{\o}egh-Krohn [2]. Unfortunately, our situation is
similar to [2] only at the first glance. In fact, there are some
essential differences. In [2] the bounded measure $d\mu$ has to
have a {\it compact}  support. Under this condition the authors
managed to apply the \textit{Kirkwood--Salsburg Equation} (KSE)
technique to prove the existence of the corresponding correlation
functions in thermodynamic limit as well as  their analyticity in
coupling constant $\l$ (in our case $\l\equiv b$) in some domain
$|\l|<\l_0$ for all temperatures.

We can not apply this technique, since of the Gaussian measure in
(1.6) has infinite support in $\R^d$. By consequence it is
impossible to find a Banach space $E_\xi$ for correlation
functions and the corresponding KSE. As it was indicated  in [2],
Remark 2, for extension of the results to the measure with
unbounded support one needs a modification of the Banach space.
But even that is not essential, since the main difficulty for $s =
1$ is connected with the parameter of analyticity. Instead of $\l$
we have as the small parameter the particle mass $m$, which is
involved in the Hamiltonian via the kinetic-energy term, see
(1.1). On the other hand it is impossible to expect the
analyticity in domain $|m|<m_*$, because we must always consider
Re$m > 0$ to define the system (1.1).

We overcome these difficulties by a \textit{cluster expansions}
technique. The choice of the cluster expansions is defined by the
dependence of the parameters in the Hamiltonian on the mass $m$
after an appropriate rescaling, see (2.6). In the case of
anharmonic potential with harmonic asymptotics $s = 1$ the
dependence on the mass $m$ remains only in the \textit{rescaled}
temperature  and in the \textit{rescaled} anharmonic potential
(1.5). Therefore, the expansions should be arranged \textit{with
respect to} (w.r.t.) the anharmonic part of the \textit{one-site
potential}. (Recall that for $s > 1$ the corresponding cluster
expansions are constructed w.r.t. two-body harmonic
\textit{interaction}.) This gives a possibility to construct the
limiting Gibbs state and to prove its analyticity in some  small
circle $|m - m_*/2| < m_*/2$ of the complex values of $m$ around
the point $m_*/2> 0$.

To prove the uniqueness of the corresponding EGM we profit the
Theorem 3.1 of [6] and we use the cluster property of the EGM
moments, which is a consequence of the cluster expansion
convergence established in the present paper. Similar to [43]  we
prove the analyticity of the state w.r.t. the external field $h$.
As a consequence this immediately implies that the displacement
\textit{order parameter} (in the region of convergence of the
cluster expansions) is equal to zero for all temperatures $T\geq
0$ as soon as $m<m_*$.

A brief contents of this paper is the following. In Section 2 we
describe our system (1.5) in more details and we formulate the
main results. In Section 3 we construct the cluster expansions
w.r.t. the anharmonic part of the one-site potential, and we give
the proof of main theorems. In Section 4 we provide all necessary
estimates.  Section 5 is devoted to the limiting Gibbs state
analyticity w.r.t. to the external field, and in Section 6 we
briefly discuss the proof of the convergence of cluster expansions
and the uniqueness in the case of high temperatures.


\section{Description of the system and main results}

We consider a $\nu (\geq 3)$-dimensional quantum crystal model
with $d (\geq 1)$-component site displacements (or $d$-dimensional
\textit{continuous infinite-spin} variables) on the cubic lattice
$ \Z^\nu $. With each site $ j \in \Z^\nu $ we associate a
one-particle physical Hilbert space $ L^{2}(\R^{d}, dq) $, where $
dq $ is the Lebesgue measure on $ \R^{d} $. Then
\begin{gather*}
\HC_{\L} \; = \; \bigotimes_{j \in \L} L^{2}(\R^{d}, dq_{j}) \; \simeq \;
L^{2}(\R^{d|\L|}, dq_{\L}), \\
dq_{\L} \; = \; \prod_{ j \in \L} dq_{j} ,
\end{gather*}
is the Hilbert space related to some given bounded set $ \L \ss
\Z^\nu$, with number of sites $ |\L|<\i $.

For a finite set $ \L \ss \Z^\nu$ we consider in $ \HC_{\L} $ the
Hamiltonian:
\begin{equation}
H_\L \; = \; H_0^\L +V^\L\label{2.1}.
\end{equation}
Here the harmonic part of the Hamiltonian has the form:
\begin{equation}
H_0^{\L} \; = \; - \frac{1}{2m}
\sum_{j \in \L} \D_j +\frac{1}{2}a
\sum_{j \in \L} q_{j}^2 + \frac{J}{4} \sum_{\langle i,j\rangle
\ss \L}(q_{i} - q_{j})^{2}+C_\L , \label{2.2}
\end{equation}
where $ m $ is the mass of particles (we put $\hbar=1$),  $\D_j$
is the Laplace operator in $L^{2}(\R^{d}, dq_{j})$, the sum is
extended over all pairs $\langle i,j\rangle\ss\L$ for which $
\|i-j\| =1 $ and $ J
> 0$. We consider a nearest-neighbors interaction only for
simplicity. The \textit{constant} $C_\L$ we will be chosen later.

We put a one-site potential $ W(q) $ in (1.1) to be equal to (1.5)
for $q^{2}= q \cdot q$, i.e., for $q\in\R^d$. So the anharmonic
part of the one-site potential in (2.1) has the form
\begin{equation}
V(q)= b ~e^{-\delta q^{2} /2},\quad b \geq 0,\delta \geq 0, \quad
q\in\R^d.\label{2.3}
\end{equation}

To prove the existence of the Gibbs states in the light-mass
regime for small temperatures, including $T=0$, we need a
supplementary argument based on the \textit{cluster expansions}.
Below we do this first for the \textit{periodic boundary
conditions} (\textit{p.b.c.}), then we complete our construction
by the case of the \textit{empty} boundary conditions The last we
use later to pass to {\it tempered non-empty} boundary conditions
Notice that the cluster expansions for \textit{p.b.c.} prove also
the clustering property of the corresponding EGM for $m<m_*$ and
for \textit{any} $T\geq 0$. Therefore, simultaneously with
existence we get \textit{uniqueness} of translation invariant EGM
for all temperature, including $T = 0$, as soon as the particle
mass is \textit{light enough}: $m<m_*$. So, first we assume in
(2.1), (2.2) the \textit{p.b.c.}:
\begin{equation*}
 \L=\{j\in\Z^\nu\mid -N_\mu/2<
j^{(\mu)}\leq N_\mu/2, \mu=1,...,\nu ; N_\mu/2\in\N\}.
\end{equation*}
Thus, $|\L|=N_1\cdots N_\nu$ and
\begin{equation}
q^{(\a)}_{(j^{(1)},...,j^{(\mu)}+N_\mu,...,j^{(\nu)})}=q_j^{(\a)},\quad
\a=1,...,d,\quad\mu=1,...,\nu.\label{2.4}
\end{equation}
Then, to complete the proof of the \textit{uniqueness} of the EGM,
in the next section we consider also a general boundary conditions
generated by some fixed configuration $\bar\om$ from the set
$\tilde\Om^t_{\hat{\b}}$, which includes so-called {\it tempered
configurations}, see [6,13,14] and  [46].

Let $ \LC(\HC_{\L}) $ be algebra of bounded operators on $
\HC_{\L} $. We  consider the temperature quantum  Gibbs state on $
\LC(\HC_{\L}) $ defined by
\begin{equation}
\r_{\L}^{\b}(A) \; = \; \frac{\Tr(A e^{-\b H_{\L}})}{Z_{\b}(\L)},
\quad \b=\frac{1}{k_B T},\label{2.5}
\end{equation}
where  $ Z_{\b}(\L) \; = \; \Tr e^{-\b H_{\L}} $
and $ A \in \LC(\HC_{\L}) $.

Note, that the algebras $ \LC(\HC_{\L}) $ (for different $ \L $)
are isometrically embedded into $ \LC(\HC_{\L'}) $ for $\L \ss \L
'$. Using this fact we can define the inductive limit
\begin{displaymath}
\AC_{0} \; = \; \lim_{\L \nearrow \Z^\nu} \LC(\HC_{\L}) ,
\end{displaymath}
which a normed $*$-- algebra of local observables. The closure of
this algebra in the operator norm forms the algebra of bounded
quasi-local observables
\begin{displaymath}
\AC \; = \; \bar{\AC}_{0}.
\end{displaymath}
The main result of the present paper is the following theorem.
\begin{Th}
For the system of quantum particles with Hamiltonian (2.1)- (2.3)
 there is  some $m_* > 0$, such that for any $0<m< m_{*}$ the
 limiting Gibbs state
\begin{displaymath}
\lim_{\L\nearrow\Z^\nu} \r_{\L}^{\b}(A) \; = \; \r^{\b}(A), \; \;
A \in \AC_{0},
\end{displaymath}
- exists  for all $T\geq 0$;\\
- it is a unique translation invariant quantum Gibbs
state (corresponding to EGM) on the algebra $ \AC_{0} $, which can
be by continuity extended to the algebra $ \AC $.
\end{Th}
Notice that by virtue of results of the paper [32](see also
[7,33,44]) it is sufficient to prove this theorem for some
sub-algebra $ \AC^{\tau} \ss \AC $ of local operators, which we
describe below. The main technical tools are the Feynman--Kac
formula and the representation of the Gibbs states (2.5) on this
algebra by functional integrals with respect to some measure $
\mu( \cdot )$ which we are going to construct.

Since we are going to consider domain of small masses, the
following standard \textit{light-mass rescaling} of variables
(appeared for the first time in this context in [58]) is relevant:
\begin{equation}
q_{j} \; = \; \a x_{j}, \; \; \a \; = \; m^{- \frac{1}{4}}.\label{2.6}
\end{equation}
This change of variables  induces the unitary map :
\begin{gather*}
U: \HC_{\L} \; \to \; \HC_{\L} \\
(Uf)(x_{\L}) \; = \; \a^{|\L|/2}f((\a x)_{\L}), \; \; (\a x)_{\L} =
\{\a x_{j}, \; j \in \L \}.
\end{gather*}
One can check that
\begin{displaymath}
U H_{\L} U^{-1} \; = \; m^{-\frac{1}{2}} \hat{H}_{\L},
\end{displaymath}
where
\begin{equation}
\hat{H}_{\L} \; = \; \hat {H}_0^\L + \hat{V}^{\L},
\label{2.7}
\end{equation}
\begin{equation}
\hat{H}_0^{\L} \; = \; - \frac{1}{2}
\sum_{j \in \L} \D_{x_j} +\frac{1}{2}a
\sum_{j \in \L}x_j^2  +
\frac{J}{4} \sum_{\langle i,j\rangle\ss{\L}}(x_{i} - x_{j})^{2}+
C_m^{\L} ,  \label{2.8}
\end{equation}
\begin{equation}
C_m^{\L}=C_{\L}m^{1/2}, \label{2.9}
\end{equation}
\begin{equation}
\hat{V}^\L \; = \;  b_m\sum_{j\in\L}e^{-\d_m x_j^2/ 2}, \quad
b_m=b m^{1/2},\quad \d_m=\d  m^{-1/2}.\label{2.10}
\end{equation}
We also define a \textit{rescaled} inverse temperature
\begin{equation}
 \hat\b=\b  m^{-\frac{1}{2}},\label{2.11}
\end{equation}
which yields
\begin{displaymath}
U \b H_{\L} U^{-1} \; = \; \hat{\b} \hat{H}_{\L}.
\end{displaymath}
Then the Gibbs state (2.5) takes the form:
\begin{displaymath}
\rho_{\L}^{\b}(A) \; = \; \hat{\rho}_{\L}^{\hat{\b}}(\hat{A}), \; \;
\hat{A} = U A U^{-1},
\end{displaymath}
where
\begin{equation}
\hat\rho_{\L}^{\hat\b}(\cdot)=
\hat{Z}_{\hat{\b}}^{-1}(\L)\Tr\left\{\cdot\exp(-\hat{\b}\hat{H}_{\L})\right\}.\label{2.12}
\end{equation}

To  represent the  state (2.12) in the form of the functional
integral we rewrite the operator $\hat{H}_0^\L$ as follows

\begin{equation}
\hat{H}_0^\L=\sum_{\a=1}^d\hat{H}_0^{(\a),\L},\label{2.13}
\end{equation}
\begin{equation}
\hat{H}_0^{(\a),\L} =-\frac{1}{2}\D_\L^{(\a)}+\frac{1}{2}
(x_\L^{(\a)},B_\L^2x_\L^{(\a)}) -\frac{1}{2}\Tr B_\L ,\label{2.14}
\end{equation}
where

\begin{equation}
\D_\L^{(\a)}=\sum_{j\in\L}\frac{\pl^2}{\pl x_j^{(\a)^2}},\quad
B_\L^2=-J\hat{\D}_{\L} + a 1\!\!1_{\L} ,\label{2.15}
\end{equation}
$x_\L^{(\a)}=(x_j^{(\a)})_{j\in\L}\in\R^{|\L|}$, and $B_\L^2$ is a
periodic $|\L|\times|\L|$ matrix, defined by the elements:

\begin{equation} \label{2.16}
 (-\hat{\D}_\L)_{jk}    =\begin{cases}
2\nu,   &\text{for }j = k, \\
   -1, &\text{for } |j-k|=1,\\
       0, &\text{otherwise },
             \end{cases}
\end{equation}
and
\begin{equation*}
(1\!\!1)_{jk}=\d_{jk}=\d_{j^{(1)}k^{(1)}}\cdots\d_{ j^{(\nu)}k^{(\nu)}  }.
\end{equation*}
Now we fix the choice of the \textit{constant} $C_\L$ in
(2.2),(2.9) by
\begin{equation}
C_\L=\frac{d}{2\sqrt{m}}\Tr B_\L.\label{2.17}
\end{equation}
By virtue of \textit{p.b.c.} the lattice operator $B_\L$ is
\textit{cyclic} (see (\ref {2.16}) and [30, Sec.16.4]), so we can
calculate its eigenvalues explicitly:

\begin{equation}
\l_k=\sqrt{\ve(k)},\quad
\ve(k)=a+4J\sum_{\mu=1}^\nu\sin^2(k^{(\mu)}/2), \quad
k\in\L^*,\label{2.18}
\end{equation}
where the set $\L^*$ is \textit{dual} to $\L$, i.e.
\begin{equation}
\L^*=\{k\in\R^\nu\mid k=\{k^{(\mu)}\}_{\mu=1}^\nu, k^{(\mu)}=
\frac{2\pi}{N_\mu}n^{(\mu)}, n^{(\mu)}=0,\pm 1,...,\pm
N_\mu-1/2,N_\mu/2\}.\label{2.19}
\end{equation}
Now the \textit{Feynman-Kac} formula (see e.g. [50],[52]) gives

\begin{equation*}
\Tr e^{-\hat{\b}\hat{H}_\L}=
\int_{\R^{d|\L|}} dx_\L\Big(e^{-\hat{\b}\hat{H}_\L}\Big)(x_\L;x_\L) =
\end{equation*}
\begin{equation}
=\int_{\R^{d|\L|}} dx_\L\int_{\Om_{\hat{\b},\L}}
W_{x_\L;x_\L}^{\hat{\b}}(d\om_\L)
e^{-\int_0^{\hat{\b}}d\tau\hat{U}(\om_\L(\tau))},\label{2.20}
\end{equation}
with
\begin{equation}
 \hat{U}(x_\L)  =\frac{1}{2}
(x_\L,B_\L^2 x_\L) -\frac{d}{2}\Tr B_\L+\hat{V}(x_\L)
,\label{2.21}
\end{equation}
where we put $ \hat{V}(x_\L)=  \hat{V}^\L $, see (2.10), and
\begin{equation*}
W_{x_\L;x_\L}^{\hat{\b}}(d\om_\L)=
\prod_{j\in\L}W_{x_j;x_j}^{\hat{\b}}(d\om_j),
\end{equation*}
here $W_{x_j;x_j}^{\hat{\b}}(d\om_j)$ is the conditional Wiener
measures on the space of continuous periodic trajectories
\begin{gather*}
\Om_{\hat{\b}, \L} \; = \; \{ \om_{\L}( \cdot ) |  \om_{\L}: S_{\hat{\b}}
\rightarrow \R^{d|\L|} \},\quad  S_{\hat{\b}}:=[0,\hat{\b}] ,\\
\om_{\L}( \cdot ) \; =
\; \{ \om_{j}( \cdot ), j \in \L | \om_{j} \in  \Om_{\hat{\b}},
\; \Om_{\hat{\b}} := C(S_{\hat{\b}} \rightarrow \R^d)\}.
\end{gather*}

Now, following [2] the \textit{unperturbed measure} in the
right-hand side of (\ref{2.20}) (i.e. $\hat{V}^\L = 0$) can be
rewritten with the help of a Gaussian integral over the Hilbert
space:
\begin{equation}
 \HC_{\hat{\b}, \L}:=\{\om_\L\in (\Om_{\hat{\b},\L},\S_{\hat{\b},\L})
\mid \|\om_\L\|^2<\infty\}.\label{2.22}
\end{equation}
Here  $ \S_{\hat{\b}, \L} $ is the standard $\s$-algebra of
$\Om_{\hat{\b},\L}$-subsets generated by Borel cylinder subsets
and norm is defined by

\begin{equation*}
\|\om_\L\|^2=\int_0^{\hat{\b}}d\tau\Big[\big(\frac{d\om_\L(\tau)}{d\tau},
\frac{d\om_\L(\tau)}{d\tau}\big)+\big(\om_\L(\tau),\om_\L(\tau)\big)\Big],
\end{equation*}
where  $(\cdot,\cdot)$ is the scalar product in $l_2(\R^{d|\L|})$.
Then for any real continuous function $F$ defined on
$\HC_{\hat{\b}, \L}$ with this unperturbed measure one gets:
\begin{equation*}
 \int_{\R^{d|\L|}} dx_\L\int_{\Om_{\hat{\b},\L}}
W_{x_\L;x_\L}^{\hat{\b}}(d\om_\L)
e^{-\frac{1}{2}\int_0^{\hat{\b}}d\tau
(\om_\L(\tau),B_{\Lambda}^2\om_\L(\tau)) +\frac{d}{2} Tr B_\L}
F(\om_\L)=
\end{equation*}
\begin{equation}
=C_0\int_{\HC_{\hat{\b}, \L}}d\mu_0^\L(\p_\L\mid p)F(\p_\L)=
C_0\int_{ \HC_{\hat{\b}, \L} } d\mu_0^\L(\p_\L)F(\p_\L)
 ,\label{2.23}
\end{equation}
where $C_0$ is a normalization constant and $d\mu_0^\L(\p_\L\mid
p)$ is the Gaussian measure  on $\HC_{\hat{\b}, \L}$ which
corresponds to \textit{p.b.c.}. We skip the index $p$ in the
following. By (\ref{2.20}) and (\ref{2.23}) it is clear that

\begin{equation}
C_0= Z_{\hat{\b}}^0(\L)=\Tr e^{-\hat{\b}\hat{H}_0^\L}.\label{2.24}
\end{equation}
Hence the right-hand side of (2.23) is the expectation:
\begin{equation}
\langle F\rangle_{\mu_0^\L}=\int_{\HC_{\hat{\b}, \L}}
d\mu_0^\L(\p_\L)F(\p_\L) \label{2.25}
\end{equation}
with respect to the Gaussian measure on the Hilbert space of
continuous periodic functions generated by the corresponding
Gaussian process with the  zero mean and the covariance $G_0^\L$:

\begin{equation*}
  G_{0;jk}^{\L;\a\a'}(\tau-\tau')=\int_{\HC_{\hat{\b}, \L}}
  d\mu_0^\L(\p_\L)
\p_j^{(\a)}(\tau)\p_k^{(\a')}(\tau')=\d_{\a\a'}
G_{0;jk}^\L(\tau-\tau').
\end{equation*}
Here $\a,\a'\in\{1,2,...,d\}$, and $G_{0;jk}^\L(\tau-\tau')$ is
the Green function of the operator\quad
$\left(-\frac{d^2}{d\tau^2}\right) 1\!\!1_\L+B_\L^2$ on
$L_2(\L\otimes S_{\hat{\b}};\R^d)$ with \textit{p.b.c.} More
explicitly [50]:

\begin{equation}
G_{0;jk}^\L(\tau-\tau')=\frac{1}{|\L|} \sum_{l\in\L^*}
e^{i(j-k)\cdot l}\quad \frac{1}{\hat{\b}}
\sum_{n\in\Z^1}\frac{\cos({2\pi}(\tau-\tau')n /
{\hat{\b}})}{({4\pi^2} n^2/{\hat{\b}^2})+ \ve(l)}\,\,,\label{2.26}
\end{equation}
or taking the sum over $n$, we get (see e.g.[50])
\begin{equation}
G_{0;jk}^\L(\tau-\tau')=\frac{1}{|\L|}\sum_{l\in\L^*}
e^{i(j-k)\cdot l}
\,\,\,\frac{e^{(\hat{\b}-|\tau-\tau'|)\sqrt{\ve(l)}}+
e^{|\tau-\tau'|\sqrt{\ve(l)}}}{2\sqrt{\ve(l)}(e^{\hat{\b}\sqrt{\ve(l)}}-1)}
\,\,.\label{2.27}
\end{equation}
It is important to note that $G_0^\L$ depends on the mass $m$ only
via \textit{rescaled} temperature $\hat{\b}=\b m^{-1/2}$.

\begin{Rk}

From (\ref{2.26})-(\ref{2.27}) one gets that for
$\L\nearrow\Z^\nu$ the covariance $G_{0;jk}^\L(\tau-\tau')$
converges pointwise to a translation invariant function
$G_{0;|j-k|}(\tau-\tau')$ which is also a bounded continuous
function and in fact it is the kernel of a positive self-adjoint
operator in $L_2(\Z^\nu\otimes S_{\hat{\b}};\R^d)$. From this
convergence it follows that the corresponding process(measure)
converges weakly to a limit process, which is a homogeneous
Gaussian process on $\Z^\nu$ and periodic on $S_{\hat{\b}}$ with
values in $\R^d$.

\end{Rk}

By virtue of (\ref{2.20}) and (\ref{2.23}) it is clear that

\begin{equation}
Z_\L=\Tr_\L e^{-\hat{\b}\hat{H}_\L}=
Z_\L^0 \int_{\HC_{\hat{\b}, \L} } d\mu_0^\L(\p)
e^{-\int_0^{\hat{\b}}d\tau\hat{V}(\p_\L(\tau))},\label{2.28}
\end{equation}
Now we can follow the line of reasoning of [2,7,32]. For every
bounded function $ \A(x_{\L}) $ on $\R^{d|\L|}$ we consider a
bounded operator $ A_{0} $ on $ \HC_{\L} $ defined as
multiplication on a bounded function:
\begin{equation}
(A_{0}f)(x_{\L}) \; = \;\A(x_{\L}) f(x_{\L}), \label{2.29}
\end{equation}
and for any $t>0$ we consider the operator
\begin{equation}
A_{t} \; = \; e^{-t\hat{H}_{\L}} A_{0} e^{t\hat{H}_{\L}}. \label{2.30}
\end{equation}
Then for every set of bounded functions $\A^{(0)},\ldots,\A^{(n)}
$ and for increasing sequence of the "time" moments $0=t_{0}\leq
t_{1}\leq\ldots\leq t_{n} \leq t_{n+1}=\hat{\b}$, we define the
operator
\begin{equation}
A \; = \; \prod_{ l=0}^n A_{t_{l}}^{(l)} \; = \; A_0^{(0)}
e^{-t_{1}\hat{H}_{\L}} A_0^{(1)} e^{-(t_{2}-t_{1})\hat{H}_{\L}}
\ldots e^{-(t_{n}-t_{n-1})\hat{H}_{\L}} A_0^{(n)}
e^{t_{n}\hat{H}_{\L}} \label{2.31}
\end{equation}
Then by (2.12) one gets:
\begin{equation}
\hat{\rho}_{\L}^{\hat{\b}}(A) \; = \;
\frac{1}{Z_{\hat{\b}}(\L)}\Tr (A_{0}^{(0)} e^{-t_{1}\hat{H}_{\L}}
A_{0}^{(1)} e^{-(t_{2}-t_{1})\hat {H}_{\L}} \ldots
e^{-(t_{n}-t_{n-1})\hat{H}_{\L}} A_{0}^{(n)}
e^{-(\b-t_{n})\hat{H}_{\L}}). \label{2.32}
\end{equation}
This definition is correct because each operator $A_{0}^{(k)}
e^{-(t_{k+1}-t_{k})\hat{H}_{\L}}$, $k = 0,1,\dots,n$, with $ t_{0}
= 0 $ and $t_{n+1}=\hat{\b}$, is of the  \textit{trace class}. The
state (2.32) is obviously analytic in the domain

\begin{equation*}
0<\mbox{Re}\,\, t_1<\mbox{Re}\,\, t_2<...< \mbox{Re}\,\,
t_n<\hat{\b}
\end{equation*}
with boundary values at $\mbox{Re}\,\, t_i=0, i=0,...,n$, and it
is uniformly bounded by the generalized Ginibre--Gruber inequality
[23]:
\begin{equation*}
\Tr\Big(\prod_{l=0}^n\a^\L_{ Im  \tau_l}(A_0^{(l)})
e^{-\hat{\b}\hat{H}_\L}\Big)\leq \|\prod_{l=0}^n
A_0^{(l)}\|\Tr\Big(e^{-\hat{\b}\hat{H}_L}\Big).
\end{equation*}
Here $\a_\tau^\L(B)$ is the $C^*$-automorphism of $\LC(\HC_\L)$:
\begin{equation*}
\a_\tau^\L(B)=e^{-i\tau \hat{H}_\L}Be^{i\tau\hat{H}_\L},\quad B\in\LC(\HC_\L)
\end{equation*}
and so the boundary value of the state $\hat{\r}_\L^{\hat{\b}}$
satisfies the KMS conditions (see [2,32]).  Moreover, the
following formula is true:

\begin{equation}
\hat{\rho}_{\L}^{\hat{\b}}( \prod_{l=0}^{n} A_{t_{l}}^{(l)} ) =
\int_{\HC_{\hat{\b},\L}} \prod_{l=0}^{n}\A^{(l)}(\p_{\L}(t_{l}))
d\mu^\L(\p) :=\langle\prod_{l=0}^{n}\A^{(l)}( \cdot
)\rangle_{\mu^\L} , \label{2.33}
\end{equation}
where $d\mu^\L$ is the \textit{Gibbsian modification} of the
measure $d\mu_0^\L$:

\begin{equation}
d\mu^{\L}(\p_\L)=Z_{\hat{\b}}(\L)^{-1}
e^{-\int_0^{\hat{\b}}d\tau\hat{V}(\p_\L(\tau))}
d\mu_0^\L(\p_\L).
\label{2.34}
\end{equation}
Following [7] we shall call this measure the \textit{Euclidean
Gibbs Measure} (EGM), which corresponds to our particular model
(2.1)-(2.3) in this context.

So, due to this construction the Theorem 2.1 for quantum Gibbs
states (2.32)-(2.33) can be reformulated as follows:

\begin{Th}
For the system of quantum particles with Hamiltonian (2.1)--(2.3)
there exists a sufficiently  small mass  $m_{*}$ such that for any
$ 0 < m <  m_{*} $ the weak limit of measures (2.34)
\begin{displaymath}
\lim_{\L \nearrow \Z^\nu} \mu^\L \; = \; \mu,
\end{displaymath}
exists and $\mu$ is a unique translation-invariant Gibbs measure
on ($\Om_{\hat{\b},\Z^\nu}$, $\S_{\hat{\b},\Z^\nu}$) in the set of
tempered Gibbs measures, for all temperatures $T\geq 0$.
\end{Th}

\begin{Rk}
To prove the existence of the limit measure we are going to apply
a cluster expansions procedure to the perturbed measure $ \mu^\L$
(2.34), which is constructed by the Hamiltonian (2.1)-(2.3) with
\textit{p.b.c.}(2.4). But, in fact, one can prove the convergence
(in the thermodynamic limit) of cluster expansions for some
non-empty set of boundary configurations of so-called tempered
type (see (3.34) of the next Section), and so to prove the
existence of the limit measure in the class of measures supported
on these tempered configurations (the set of {\it tempered Gibbs
measures}).

To prove the uniqueness of the limit measure in this set we use
one elegant criterium of uniqueness due to [6] (Theorem 3.1).
\end{Rk}

\begin{Rk}
To prove the existence of the limit measure $\mu$  it is
sufficient to prove the convergence of the averages (2.33) for
local observables

\begin{equation}
\A_B(\p_B)=\prod_{j\in B}\A_j(\p_{\tau_j})\label{2.35}
\end{equation}
for any bounded subsets $B\ss\L$ and $\tau_j\in S_{\hat{\b}}$.
\end{Rk}

\begin{Rk}
Theorem 2.1  is equivalent to the Theorem 2.2  on the subalgebra
constructed by (2.29)-(2.31). Then we use the fact that any
operator $A\in\AC$ may be approximated  by a linear combination of
operators of the form (2.31) (see Theorem 2.2 of [32]). Hence the
state $\hat{\r}_\L^{\hat{\b}}(A)$ will also be approximated by
(2.33).
\end{Rk}

\section{Cluster expansions and proof of main Theorems}

\begin{Rk}
Before we construct the cluster expansions, which we need  for the
proof of Theorem 2.2 we recall the difference between the case
$s>1$ in (1.3) and our case (1.5) with harmonic growth at infinity
(see Section 1 and discussion in [43,44]). Notice that for
harmonic interaction (i.e. for polynomials of the second degree)
the rescaling (2.6) implies, for $s>1$, a week interaction between
oscillators proportional to $m^{(s-1)/(s+1)}$. But for the
harmonic growth of the one-site potential and for harmonic
interaction the rescaling (2.6) leads to smallness of anharmonic
perturbation $\hat{V}^\Lambda$, see (\ref{2.10}).
\end{Rk}

\begin{Rk}
Notice that the cluster expansions developed in [8,36,44] for
$s>1$ are suited for small interaction between oscillators. In
contrast to that the cluster expansions presented in the present
paper are relevant for small one-site anharmonism and harmonic
oscillators interaction. Therefore, it is suitable to arrange our
reference measure in such a way it includes this interaction.
Following the paper [2] it gives us the Gaussian measure
representation for the state.
\end{Rk}

The cluster expansions technique for the Gaussian integrals with
non-diagonal covariance was developed in [29] (see also [30,49]).
Then it was applied in [20,22] for the proof of the Debye
screening in dilute charged particle systems. This type of cluster
expansions includes derivatives acting on the Gibbs factor
$\exp[-\hat{V}^\Lambda]$. Since the coefficient $b_m$ is
proportional to $m^{1/2}$, and the coefficient $\d_m$ is
proportional to $m^{-1/2}$ (see (\ref{2.10})), we have no
analyticity at $m=0$. Therefore, the estimates are delicate in the
small-mass domain.

To prove the convergence of the measure $d\mu^\L$ in the
thermodynamic limit it is sufficient to prove the existence of
limit state for observables
 $\A_B$ (2.35).

We would like to notice that similar to non-polynomial quantum
field theory (see e.g. [30,49]) our cluster expansions correspond
to a one-site perturbation, and not to the interaction between
sites. To start the construction of cluster expansions we define
new configuration space. First we consider $S_{\hat{\b}}$ for
large $\hat{\b}$ including $\hat{\b}=\infty$. So for a fixed $m$
we consider sufficiently low temperature such that critical
behavior (if any) might be in the interval $[0,\hat{\b}]$. Then
just for convenience we consider $\hat{\b}\in\N$ to define the
partition of the interval $[0,\hat{\b}]$ into \textit{unit}
intervals. So, we define

\begin{equation}
\bar\D_{\hat{\b}} \; := \; \Big\{ \D_{\hat{\tau}}
:=[\hat{\tau},\hat{\tau}+1]\mid \hat{\tau}=0,1,...,\hat{\b}-1,
\quad \hat{\b}\in\N\Big\},\label{3.1}
\end{equation}
Then we define
\begin{equation}
t:=(j,\tau),\quad j\in\Z^\nu,\quad \tau\in
S_{\hat{\b}},\label{3.2}
\end{equation}
and the  box
\begin{equation}
T:=\L\otimes S_{\hat{\b}}\ss \Z^\nu\otimes\R_+,\label{3.3}
\end{equation}
where $\L$ and $S_{\hat{\b}}$ are defined in the previous section.
With these notations we introduce
\begin{equation}
G_{0}^{T}(t, t'):= G_{0;jj'}^\L(\tau,\tau').\label{3.4}
\end{equation}
For any $j\in\Z^\nu$ and $\D_{\hat{\tau}}\in\bar\D_{\hat{\b}}$ we put
\begin{equation}
\D_{j,\hat{\tau}}:=(j,\D_{\hat{\tau}}),\label{3.5}
\end{equation}
and consider $T$ as finite union of "rods" $\D_{j,\hat{\tau}},
j\in\Z^\nu,\quad \hat{\tau} = 0,1,...,\hat{\b}-1$.

To construct the cluster expansions for the expectations of
(\ref{2.35}) we define (following [20]) a family of sets by the
\textit{inductive procedure}. Let for $n=1$
\begin{equation}
X_1=Y_1=(B,\D_B),\quad \D_B=\bigcup_{j\in B}\D_j,\label{3.6}
\end{equation}
where  $\D_j$ coincides with   $\D_{\hat{\tau}}$ such that
$\tau_j\in\D_{\hat{\tau}}$. For $n\geq 2$ we put
\begin{equation}
Y_n=(j,\D_{\hat{\tau}}), j\in\L\sm B,\quad \text{or} \quad
\D_{\hat{\tau}}\in\bar{\D}_{\hat{\b}}\sm\D_B, \label{3.7}
\end{equation}
and define
\begin{equation}
X_n :=X_{n-1}\bigcup Y_n,\quad X_n^c:=T\sm X_n.\label{3.8}
\end{equation}
So, $Y_n$ for $n\geq 2$ are "rods" $\D_{j,\hat{\tau}}$, and $X_n$
are their unions.

Now define the sequence of new covariances (for $0\leq s_i\leq 1,
i=1,...,n$):
\begin{equation}
G_0^T(t,t';(s)_n)= G_0^T(t,t')p(t-t';(s)_n),\label{3.9}
\end{equation}
where
\begin{equation*}
p(t,t';(s)_n)=\sum_{m=1}^{n+1} \1_{Y_m}(t)\1_{Y_m}(t')+
\end{equation*}
\begin{equation}
\sum_{1\leq l<m\leq n+1}s_ls_{l+1}\cdots s_{m-1} \big[
\1_{Y_l}(t)\1_{Y_m}(t')+ \1_{Y_m}(t)\1_{Y_l}(t')\big].
\label{3.10}
\end{equation}
Here $\1_Y(t)$ is indicator of the ``rod'' $Y$, and the ``rod''
$Y_{n+1}$ we identify with $X_n^c$.

One can treat  $G_0^T(t,t';(s)_n)$ as a kind of ``interaction
potential'' in the space of "cluster" configurations
$\L\times\bar\D_{\hat{\b}}$. Then the interpolation parameters
$s_i$ specify the intensity of this ``interaction'' between "rods"
of $X_i$ and of $X_i^c$. Putting $s_n=0$ we get that there is no
``interaction'' between the ``rods'' of $X_n$ and ``rods'' of
$X_n^c$. Then let $\mu^T_{0;(s)_n}$ be the Gaussian measure with
covariance $G_0^T(t,t';(s)_n)$. This is the kernel of a
positive-definite operator in $L_{2}(T;\R^d)$ because it is a
linear combination with positive coefficients of operators of the
form $\1_X G_0^T\1_X$, where $X$ is some of the unions of $Y_m$,
see [21], [30, Sec.18.2] and Section 4 for details. It is clear
from (3.10) that $\mu^T_{0;(1)_n}=\mu_0^\L$. For $s_n=0$ one gets
that the measure $d\mu^T_{0;(s)_{n-1},0}$ becomes the product
measure
\begin{equation}
d\mu^T_{0;(s)_{n-1},0}=d\mu^{\L,X_n}_{0;(s)_{n-1}}
d\mu^{\L,X_n^c}_{0;0},\label{3.11}
\end{equation}
where the covariance of $d\mu^{\L,X_n}_{0;(s)_{n-1}}$ is

\begin{equation}
G^{\L,X_n}_0(t,t';(s)_{n-1},0)= G_0^T(t,t';(s)_{n-1},0)-
\1_{X_n^c}(t)G_0^T(t,t')\1_{X_n^c}(t'),\label{3.12}
\end{equation}
and the covariance of $d\mu^{\L,X_n^c}_{0;0}$ is

\begin{equation}
G^{\L,X_n}_0(t,t')=
\1_{X_n^c}(t)G_0^T(t,t')\1_{X_n^c}(t').\label{3.13}
\end{equation}
Notice, also, that $(s)_{n-1}\Big|_{n=1}=\emptyset$ and
$d\mu^{\L,X_n}_{0;(s)_{n-1}}\Big|_{n=1}=:d\mu^{\L,X_1}_0$, and
that each step in the expansion can be obtained by applying the
Newton-Leibnitz formula:
\begin{equation*}
\langle\A_B( \cdot )\rangle_{\mu^{\L}}=Z_T^{-1}
\int_{\HC_{\hat{\b},\L}}  d\mu_{0;0}^T(\p)\A_B(\p_B)e^{-\hat{V}(\p_T)}+
\end{equation*}
\begin{equation*}
Z_T^{-1}\int_0^1 ds_1\frac{d}{ds_1} \int_{\HC_{\hat{\b},\L}}
d\mu_{0;s_1}^T(\p)\A_B(\p_B)
e^{-\hat{V}(\p_T)}=
\end{equation*}
\begin{equation}
 \int_{\HC_{\hat{\b},\L}}  d\mu_{0}^{\L,X_1}(\p_{X_1})\A_B(\p_B)
e^{-\hat{V}(\p_{X_1})} \frac{Z_T(X_1^c)}{Z_T} +R_1^\L,
\label{3.14}
\end{equation}
where $Z_T:=Z_{\hat{\b}}(\L)$ and
\begin{equation}
Z_T(X)= \int_{\HC_{\hat{\b},\L}}  d\mu_{0}^T(\p)
e^{-\hat{V}(\p_X)}.\label{3.15}
\end{equation}
Using integration by parts formula to calculate the derivative
w.r.t. parameter $s_1$ (see, for example, [30, Sec.9.1]) we obtain

\begin{equation}
R_1^\L=Z_T^{-1}\int_0^1 ds_1 \int_{\HC_{\hat{\b},\L}}
d\mu_{0;s_1}^T(\p)\D_\p\A_B(\p_B)e^{-\hat{V}(\p_T)},\label{3.16}
\end{equation}
where
\begin{equation}
\D_\p=\sum_{\a=1}^d\D_\p^{(\a)},\label{3.17}
\end{equation}

\begin{equation*}
\D_\p^{(\a)}=\frac{1}{2}\sum_{j,k\in\L}\int_0^{\hat{\b}}d\tau
\int_0^{\hat{\b}}d\tau' G_{0;jk}^\L(\tau-\tau')
\frac{d}{ds_1}p\big((j,\tau),(k,\tau');s_1\big)
\frac{\d^2}{\d\p_j^{(\a)}(\tau)\d\p_k^{(\a)}(\tau')}
\end{equation*}

\begin{equation*}
=: \frac{1}{2} \int_Tdt
\int_Tdt' G_{0}^T(t,t')
\frac{d}{ds_1}p(t,t';s_1)
\frac{\d^2}{\d\p^{(\a)}(t)\d\p^{(\a)}(t')}
\end{equation*}

\begin{equation*}
=\int_{Y_1}dt
\int_{X_1^c}dt' G_{0}^T(t,t')
\big(\frac{d}{ds_1}s_1\big)
\frac{\d^2}{\d\p^{(\a)}(t)\d\p^{(\a)}(t')}
\end{equation*}

\begin{equation}
=\sum_{Y_2\in X_1^c} \big(\frac{d}{ds_1}s_1\big)
\D_{1,2}^{(\a)}.\label{3.18}
\end{equation}
For  definition of $\D_{p,p'}^{(\a)}$  ($p, p'\in\N$) see (3.24).
Then, taking into account that

 \begin{equation}
e^{-\hat{V}(\p_T)}= e^{-\hat{V}(\p_{X_2})}e^{-\hat{V}(\p_{X_2^c})}
\label{3.19}
\end{equation}
with $X_2=X_1\cup Y_2= Y_1\cup Y_2$, and putting (3.17)-(3.19) in
to (3.16) we repeat the procedure in each term of the sum over
$Y_2$ to carry out the second step of expansion. After $n_T=|T\sm
X_1|$ steps all "rods" in $T$ will be exhausted and we get the
identity:

\begin{equation}
\langle\A_B( \cdot )\rangle_{\mu^{\L}}= \sum_{n=1}^{n_T}\sum_{\bar
Y\ss T} \KC^T_{\bar Y}(\A_B) F^T_n(\bar Y),\label{3.20}
\end{equation}
where
\begin{equation}
\KC^T_{\bar Y}(\A_B)= \int(ds)^{n-1} f(\eta;(s)_{n-2})I^T_n(\bar
Y,\eta,\A_B;(s)_{n-1}),\label{3.21}
\end{equation}

\begin{equation}
I^T_n(\bar Y,\eta,\A_B;(s)_{n-1})=\int_{\HC_{\hat{\b},\L}}
d\mu^{\L,X_n}_{0;(s)_{n-1}}(\p_{X_n}) \D(\eta,\bar
Y)\A_B(\p_B)e^{-\hat{V}(\p_{X_n})},\label{3.22}
\end{equation}
with
\begin{equation}
\D(\eta,\bar Y)=\prod_{l=2}^n\D_{\eta(l),l},\quad \D_{p,p'}=
\sum_{\a=1}^d \D_{p,p'}^{(\a)},\label{3.23}
\end{equation}
where $\eta :=\{(\eta(l),l)_{l=2}^n\}$ is a sequence of couples
with $\eta(l)<l$, which we use to construct the \textit{trees}
estimates according the standard procedure, see e.g. [21], or [42]
Ch.II.4. Here
\begin{equation}
\D_{p,p'}^{(\a)}=\int_{Y_p}dt\int_{Y_{p'}}dt^{'}G_0^T(t,t^{'})
\frac{\d^2}{\d\p^{(\a)}(t)\d\p^{(\a)}(t^{'})},\label{3.24}
\end{equation}

\begin{gather}
f(\eta;(s)_{n-2})
=\prod_{2\leq m\leq n}(s_{\eta(m)}s_{\eta(m)+1}
  \cdots s_{m-2}), \label{3.25} \\
 \eta(m)\leq m-2,\quad s = (s_{1},\ldots, s_{n-1}),\quad
 0 \leq s_{j} \leq 1.\label{3.26}
\end{gather}
The second sum in (3.20) is over all possible sets $\bar
Y=\{Y_2,...,Y_n\}$ in $T$. The product in (3.25) is equal to 1 if
$ \eta(m) = m-1 $. Finally

\begin{equation}
F^T_n(\bar Y)=\frac{Z_T( X_n^c)}{Z_T}, \quad Z_T(X_n^c)=Z_{T}(T\sm
X_n).\label{3.27}
\end{equation}
(For more details about construction of this type of expansions
see, for example the survey [49].)

To prove Theorem 2.2 we use the representation (3.20) and the
following collection of lemmas.
\begin{Lm}
With the same assumptions as in Theorem 2.2
 there exists a sufficiently small value of
the mass $m_{*}$,
and constant $C_B$ which does not depend on $T$ and on $m$,
such that for all $0<m<m_{*}$
\begin{equation}
\sum_{\bar Y\ss \Z^\nu\otimes S_{\hat{\b}}}\KC^T_{\bar Y}(\A_B)
\le C_B\ve^{n-1},\label{3.28}
\end{equation}
where  $\ve=\ve(m)\to 0$ as $m\to 0$ and
\begin{equation}
\lim_{\L \nearrow \Z^\nu} \KC^T_{\bar Y}(\A_B)=\KC^{\b}_{\bar
Y}(\A_B).\label{3.29}
\end{equation}
\end{Lm}

\begin{Lm}
With the same assumptions as in Theorem 2.2 there exists a constant
$c$  such that
\begin{equation}
 F^T_{n}(\bar Y)
  \le e^{c(|B|+n-1)}.\label{3.30}
\end{equation}
The constant $c$ does not depend on $\L, \hat{\b}$, $B$, $m$,  and
one has the limit:
\begin{equation}
\lim_{\L \nearrow \Z^\nu} F^T_{n}(\bar Y)=F_n^{\hat{\b}}(\bar
Y).\label{3.31}
\end{equation}
\end{Lm}
We also need the following corollary of  Lemmas 3.1 and 3.2.

\begin{Co}

Let $B'\ss\L$ and $B\cap B'=\emptyset$
and let the sum over $\bar Y$ be restricted so
that $X_n\cap B\ne\emptyset$. Then

\begin{equation}
\sum_{n=1}^\infty\sum_{\bar Y, X_n\cap B'\ne\emptyset}|\KC^T_{\bar Y}(\A_B)|
 F^T_{n}(\bar Y)
\le C'_B e^{-\sqrt{a}\mbox{dist}(B,B')},\label{3.32}
\end{equation}
where $a$ is the constant of the one-site  harmonic potential.
\end{Co}

Now, using the standard ``doubling  measure''   trick by J.
Ginibre [28] (or [30, Sec.18.3])
 we obtain from Corollary 3.1 the exponential clustering of
correlations:

\begin{Lm}
With the same assumptions as in Theorem 2.2 and for any measurable
functions $\A_{B_1}$,$\A_{B_2}$ w.r.t. $\s$-algebras
 $ \S_{\hat{\b},B_1} $ and $ \S_{\hat{\b},B_2} $ respectively, where
$B_1, B_2\ss \L$ and $B_1\cap B_2=\emptyset$, there exists a
sufficiently small value of $m_*$, such that for $0< m < m_*$ and
any temperature one has the estimate

\begin{equation*}
\langle\A_{B_1}( \cdot ) \A_{B_2}( \cdot )\rangle_{\mu^{\L}}-
\langle\A_{B_1}( \cdot )\rangle_{\mu^{\L}}
\langle \A_{B_2}( \cdot )\rangle_{\mu^{\L}}\leq
\end{equation*}
\begin{equation}
\leq C_1(\A_{B_1})
C_2(\A_{B_2})e^{-\sqrt{a}\mbox{dist}(B_1,B_2)}.\label{3.33}
\end{equation}
uniformly in $\L$.
\end{Lm}

The proofs of  Lemmas 3.1-3.2 and Corollary 3.1 are presented in
the next Section. They imply the proof of our Theorem 2.2 for the
case of \textit{p.b.c.}

To prove the uniqueness of the limit measure in the Theorem 2.2
for \textit{non-periodic} boundary conditions, we keep the DLR
language (see [25]) and consider EGM, with some general
\textit{boundary conditions} $\xi$, from a class studied already
in  [6,7]. To this end we define the set of {\it tempered}
configurations:

\begin{equation}
\Tilde\Om^t_{\hat{\b}}:=\bigcup_{\r>0} \tilde\Om^\r_{\hat{\b}}=
\bigcup_{\r>0}\left\{\xi\in\Om_{\hat{\b},\Z^\nu}\mid \,\,
\sum_{l\in\Z^\nu} e^{-\r|l|}\,
\|\xi_l\|_{L^2[0,\hat{\b}]}<\infty\right\}.\label{3.34}
\end{equation}

\begin{Rk}
For tempered configurations $\Om^t_{\hat{\b}}$ defined in [6] one
has:  $\Om^t_{\hat{\b}} \ss \Tilde\Om^{t}_{\hat{\b}}$,  since we
include also configurations that may have an  exponential growth
at infinity.
\end{Rk}

First we define the harmonic measure with \textit{zero } boundary
conditions $\mu_0^\L(\cdot\mid 0)$ in the same way as it was done
in (2.23) for $\mu_0^\L(\cdot\mid p)$. Then for any
$\xi\in\tilde\Om^\r_{\hat{\b}}$ with $\r<\sqrt{a}$, where $a$ is
the one-site harmonic constant from (2.8), we define the perturbed
measure with \textit{non-zero } boundary conditions by

\begin{equation}
d\mu^\L(\p\mid\xi)\; := \; Z_{\hat{\b}}(\L\mid\xi)^{-1}
e^{-\int_0^{\hat{\b}}d\tau\hat{V}(\p_\L(\tau))+
\frac{J}{2}\sum_{\langle
l,l'\rangle}^{\pl\L}\int_0^{\hat{\b}}d\tau
\p_l(\tau)\xi_{l'}(\tau)} d\mu_0^\L(\p\mid 0),\label{3.35}
\end{equation}
where

\begin{equation}
\sum_{\langle l,l'\rangle}^{\pl\L}(...)
:=\sum_{l\in\L,l'\in\L^c,|l-l'|=1}(...).\label{3.36}
\end{equation}
We also need the following \textit{auxiliary measure}
$\tilde\mu^\L(\cdot\mid y_\L)$ on $\HC_{\hat{\b},\L}$, which
depends on some fixed trajectories $y_\L\in\Om_{\hat{\b},\L}$:

\begin{equation}
d\tilde\mu^\L(x_\L\mid y_\L):=\tilde Z_{\hat{\b}}(\L\mid
y_\L)^{-1} e^{-\int_0^{\hat{\b}}d\tau\Tilde V(x_\L(\tau)\mid
y_\L(\tau))} d\mu_0^\L(x_\L\mid 0),\label{3.37}
\end{equation}
where (see (2.10)):

\begin{equation}
\Tilde V(x_\L(\tau)\mid y_\L(\tau))= b_m\sum_{j\in\L}\Big[e^{-
\d_m(x_j(\tau)+y_j(\tau))^2 /4}+ e^{- \d_m(x_j(\tau)-y_j(\tau))^2
/4}\Big].\label{3.38}
\end{equation}
It is clear that with the measure $\tilde\mu^\L(\cdot\mid y_\L)$
one can construct the \textit{same} cluster expansion as in
(3.23), but with \textit{additional} dependence on  configurations
$y_\L$:

\begin{equation}
\langle\A_B( \cdot )\rangle_{\tilde\mu^{\L}(\cdot\mid y_\L)}=
\sum_{n=1}^{n_T}\sum_{\bar Y\ss T} \KC^T_{\bar Y}(\A_B\mid y_\L)
F^T_n(\bar Y\mid y_\L),\label{3.39}
\end{equation}
In the next Section we prove the following lemma:

\begin{Lm}
Let $m <  m_*$ as in Lemmas 3.1-3.3. Then for the measure
$\tilde\mu^\L(\cdot\mid y_\L)$, defined by (3.37)-(3.38), the
estimates (3.30)-(3.33) are true with the same constants which do
not depend on configurations $y_\L$.
\end{Lm}
{\bf Proof of Theorem 2.2.} \\
{\it Existence.} First we prove the existence of measures with
\textit{p.b.c.} and \textit{zero } boundary conditions As we
mentioned in Remark 2.1 the covariance $G_{0;jk}^\L(\tau-\tau')$
converges pointwise to the translation invariant function
$G_{0;|j-k|}(\tau-\tau')$ and the corresponding measures $d\mu_0$
and $d\mu_{0;(s)_{(n-1)}}^{X_n}$ are defined on
 $\S_{\hat{\b}}$. So, using Lemmas 3.1 and 3.2 we can rewrite the
expansion (3.20) in the \textit{thermodynamic limit}:
\begin{equation}
\langle\A_B( \cdot )\rangle_{\mu}= \sum_{n\geq 1}\sum_{\bar Y\ss
\Z^\nu\otimes S_{\hat{\b}}} \KC^{\hat{\b}}_{\bar Y}(\A_B)
F^{\hat{\b}}_n(\bar Y),\label{3.40}
\end{equation}
with the same definitions of $I_n$ and $\D(\eta,\bar y)$ (see
(3.22), (3.23)),  but with $G_0$ and $\mu_{0;(s)_{n-1}}^{X_n}$
instead of $G_0^T$ and $\mu_{0;(s)_{n-1}}^{\L,X_n}$.

 Hence, there exists  probability measures $\mu_{B}$ indexed by
$B\ss\Z^\nu$ such that
 \begin{displaymath}
  \langle\A_B\rangle=\int_{\HC_{\hat{\b},\L}} \A_B(\p)\,d\mu_{B}(\p).
 \end{displaymath}
All these measures are consistent and consequently by the
Kolmogorov Theorem one can construct a unique limit measure $\mu$
on the $\s$-algebra $\Sigma_{\hat{\b},\Z^\nu}$. \hfill \hfill
$\square$
\begin{Rk}
In the case of general boundary conditions
$\xi\in\Tilde\Om^\r_{\hat{\b}}$ with $\r<\sqrt{a}$, the factors
$\KC^{\hat{\b}}_{\bar Y}$, $F^{\hat{\b}}_n$, and the constants
$c$, $C_B$ in the Lemmas 3.1-3.2 will depend on $\xi$. This means
that from convergence of the cluster expansions we can obtain only
existence of the limit
measure, but not uniqueness.\\
\end{Rk}
{\it Uniqueness.} To prove the uniqueness we use the following
\begin{Pn}(see Theorem 3.1 of [6])\\
Suppose that for every $l_0\in\Z^\nu$ and $\tau_0\in
[0,\hat{\b}]$, for every increasing sequence of boxes
$\L_n\ss\Z^\nu$ ($\L_n\rightarrow\Z^\nu$), such that $l_0\in\L_n$
for all $n$, and for any two configurations $\xi,
\eta\in\tilde\Om^\r_{\hat{\b}}$, $\r<\sqrt{a}$, one has the
convergence:

\begin{equation}
\langle\p_{l_0}( \tau_0 )\rangle_{\mu^{\L_n}(\cdot\mid\xi)}-
\langle\p_{l_0}( \tau_0 )\rangle_{\mu^{\L_n}(\cdot\mid\eta)}
\rightarrow 0 \label{3.41}
\end{equation}
as $n\rightarrow\infty$. Then there exist a unique limit measure
$\mu$ which does not depend on configurations $\xi, \eta$.

\end{Pn}

This statement is a quantum version of some known previous
results, namely, the monotonicity arguments [41]  for the measures
on the configurations in the Classical Statistical Mechanics and a
priori estimates for EGM due to  [10,11]. These arguments are true
for the wide class of anharmonic potentials with
\textit{ferromagnetic} type of interaction. We use this statement
as follows. Consider

\begin{equation*}
\langle\p_{l_0}( \tau_0 )\rangle_{\mu^{\L_n}(\cdot\mid\xi)}-
\langle\p_{l_0}( \tau_0 )\rangle_{\mu^{\L_n}(\cdot\mid\eta)} =
\end{equation*}
\begin{equation}
=\int_0^1ds \frac{d}{ds}\langle\p_{l_0}( \tau_0
)\rangle_{\mu^{\L_n}(\cdot\mid\eta+ s\zeta)},\quad
\zeta=\xi-\eta.\label{3.42}
\end{equation}
After differentiation we obtain:

\begin{equation*}
|\langle\p_{l_0}( \tau_0 )\rangle_{\mu^{\L_n}(\cdot\mid\xi)}-
\langle\p_{l_0}( \tau_0 )\rangle_{\mu^{\L_n}(\cdot\mid\eta)}|\leq
\end{equation*}
\begin{equation}
\sup_{s\in [0,1]}\frac{J}{2}\sum_{\langle l,l'\rangle}^{\pl\L}
\int_0^{\hat{\b}} d\tau K^\L_{ll_0}(\tau,\tau_0\mid\eta+
s\zeta)\zeta_{l'}(\tau),\label{3.43}
\end{equation}
where

\begin{equation}
 K^\L_{ll'}(\tau,\tau'\mid\eta)=
\langle\p_{l}( \tau )\p_{l'}( \tau' )
\rangle_{\mu^{\L}(\cdot\mid\eta)}- \langle\p_{l}( \tau
)\rangle_{\mu^{\L}(\cdot\mid\eta)} \langle\p_{l'}( \tau'
)\rangle_{\mu^{\L}(\cdot\mid\eta)}.\label{3.44}
\end{equation}
It is easy to see that the convergence (3.41) follows from the

\begin{Lm}
For $m< m_*$ there are constants $C_i, i=1,2$ independent of
 $\tau, \tau', \L, \eta$, such that

\begin{equation}
 K^\L_{ll'}(i)\leq C_i\; e^{-\sqrt{a}|l-l'|},\label{3.45}
\end{equation}
where
\begin{equation}
 K^\L_{ll'}(1)=K^\L_{ll'}(\tau,\tau'\mid\eta),\quad \mbox{and}
\quad K^\L_{ll'}(2)=\int_0^{\hat{\b}}d\tau'
K^\L_{ll'}(\tau,\tau'\mid\eta).\label{3.46}
\end{equation}
\end{Lm}
Finally, applying (similar to [6]) the Cauchy-Schwarz inequality
in (3.43) and using (3.45), we get (3.41).  \qquad \hfill $\square$\\
{\bf Proof of Theorem 2.1.}\\
As we mentioned in Section 2, our Theorem 2.2 is, in fact, a
reformulation of Theorem 2.1 on the algebra which is constructed
by (2.29)-(2.33). The uniqueness of the limit Gibbs state on the
corresponding algebra follows from (3.43)-(3.46) by virtue of
Proposition 3.1 for the
$(\Om_{\hat{\b},\Z^\nu},\S_{\hat{\b},\Z^\nu})$ on the set of
tempered configurations (3.34). The convergence of cluster
expansions gives also \textit{analyticity} of the limit state in
the circle $|m-m_*/2|<m_*/2$ of the complex plane $m$. The end of
the proof is based on the same arguments as in Theorem 4.1 of [2].
 \qquad \hfill $\square$


\section{Convergence of the cluster expansions}

To prove the convergence of the cluster expansions (3.20) for
$\L\nearrow\Z^\nu$ we should first give demonstrations of Lemmas 3.1 and 3.2.\\
{\bf Proof of Lemma 3.1.} To simplify the proof we omit indexes
$\a=1,...,d$ and we take into account that  the number of
derivatives for each order of $n$ should increase as $d^{n-1}$. To
write down the general term for any fixed \textit{tree}  $\eta$
(see (3.23)) let us introduce (following [16,17]) the numbers
$d_\eta(k)$, $k=1,...,n$ which are the characteristics of tree
$\eta$, i.e. $d_\eta(k)$ is a number of lines, coming into the
vertex $k$ from vertices $m$ $(m > k)$. It is clear that
$n_1=d_\eta(1)$, $n_k=d_\eta(k)+1$ for $k=2,...,n$, are exactly
the number of lines coming from vertices $k=1,...,n$ and these
numbers coincides with the numbers of derivatives supported in
$Y_1,Y_2,...,Y_n$. Hence,

\begin{equation}
\sum_{k=1}^n d_\eta(k)=n-1\quad \mbox{and}
\quad \sum_{k=1}^n n_k=2(n-1).\label{4.1}
\end{equation}
Let us rewrite the operator $\D(\eta,\bar Y)$ in the following
form
\begin{equation*}
\D(\eta,\bar
Y)=\Big(\prod_{k=1}^n\prod_{l=1}^{n_k}\int_{Y_k}dt^{(l)}_k\Big)
\prod_{k=2}^n G_0^T(t_{\eta(k)}^{(\eta_k)}-t_k^{(n_k)})\times
\end{equation*}
\begin{equation}
\prod_{k=1}^n\frac{\d^{n_k}}{\d\p(t_k^{(1)})\cdots\d\p(t_k^{(n_k)})}.
\label{4.2}
\end{equation}
Here we put
\begin{equation*}
\eta_2:=1,\quad
\eta_k:=1+\sum_{l=2}^{k-1}\d_{\eta(l),\eta(k)},\quad k\geq 3,
\end{equation*}
where $\d_{\eta(l),\eta(k)}$  is the symbol of Kronecker. In the
following we also put $t_k^{(n_k)}:=t_k$.

The main technical estimate is the following
\begin{Pn}
Let $D_{n_k}$ be parts of operator  $\D(\eta,\bar Y)$ and factor
$e^{-\hat{V}(\p_{X_n})}$ corresponding to the $k$-th vertex on the
tree $\eta$:
\begin{equation*}
D_{n_k}=D_{n_k}(t_{\eta(k)}^{(\eta_k)};t_1,...,t_{d_\eta(k)})=
\end{equation*}
\begin{equation*}
=\int_{(Y_k)^{n_k}}G_0^T(t_{\eta(k)}^{(\eta_k)}-t_k^{(n_k)})
G_0^T(t_{k}^{(1)}-t_1)\cdots G_0^T(t_{k}^{(d_\eta(k))}-t_{d_\eta(k)})
\times
\end{equation*}
\begin{equation}
\times\frac{\d^{n_k}}{\d\p(t_k^{(1)})\cdots\d\p(t_k^{(n_k)})}
e^{-\hat{V}(\p_{Y_k})}. \label{4.3}
\end{equation}
Then for $b_m < 1$ the following estimate is true:
\begin{equation*}
D_{n_k}(t_{\eta(k)}^{(\eta_k)};t_1,...,t_{d_\eta(k)})\leq
2^{n_k}b_m \d_m^{n_k/2} n_k!\times
\end{equation*}
\begin{equation}
\times\sup_{(*)_{d_\eta(k)}}\int_{Y_k}dt_k
G_0^T(t_{\eta(k)}^{(\eta_k)}-t_k^{(n_k)})
G_0^{*,T}(\cdot-t_1)\cdots
G_0^{*,T}(\cdot-t_{d_\eta(k)})e^{-\d_m\p(t_k)^2/ 4 },\label{4.4}
\end{equation}
here supremum is taken over all terms in the product, where one
can choose $G_0^{*,T}(\cdot-t_i)$ equal either to
$G_0^{T}(t_k-t_i)$, or $\int_{Y_k}dt'G_0^{T}(t'-t_i)$.
\end{Pn}
{\it Proof.} For simplicity  we demonstrate this proposition for
the case when instead of functional derivatives we have the
ordinary derivatives, since exactly the same line of reasoning
gives the proof for functional derivatives.

Let $X=-b_me^{-\d_mx^2/ 2}$ denote the "scalar prototype" of our
anharmonic potential in (4.3), supported in $Y_k$. Then it easy to
estimate by induction the value of $n$-th derivatives of $X$:
\begin{equation}
|X^{(n)}|\leq b_m2^n\d_m^{n/2}\sqrt{n!}e^{-\d_mx^2/ 4
}.\label{4.5}
\end{equation}
This inequality can be obtained by induction using the recursion relation:
\begin{equation}
I_n(x)=-\d_mxI_{n-1}(x)-(n-1)\d_mI_{n-2}(x),\quad
I_n(x)=e^{\d_mx^2 /4}X^{(n)}(x),\label{4.6}
\end{equation}
and the inequality:
\begin{equation*}
|x|^ke^{- \d_mx^2 /4}\leq 2^{k/2}\d_m^{-k/2}\sqrt{k!} ,\,.
\end{equation*}
Now using the Leibnitz formula for the product $X^{(1)}e^{X}$,
where the first derivative $X^{(1)}=b_m\d_mxe^{-\d_mx^2/ 2}$, and
the estimate (4.5), again by induction one obtains:
\begin{equation}
|\frac{d^n}{dx^n}e^{X}|\leq 2^nb_m\d_m^{n/2}n! e^{-\d_mx^2/ 4}.
\label{4.7}
\end{equation}
We also used here the fact that $e^X\leq 1$.
\begin{Rk}
The inequality (4.7) explains the provenance of the coefficient in
(4.4), since by the same way this coefficient appears when one
passes to functional derivatives (as in (4.3)), instead of the
ordinary ones.
\end{Rk}

\begin{Rk}
Notice that we also should consider the terms in which some
derivatives act on the factor $\A_B$.  But, since $\A_B$ is
supported only in $Y_1=X_1$, then even in the case, when for a
given tree $\eta$ one has all $\eta(k)=1$, for $k=2,...,n$, and
$(n-1)$ derivatives act only on $\A_B$, the other $(n-1)$
derivatives in $\D(\eta,\bar Y)$ act on the factor
$\exp[-\hat{V}_a(\p_{X_n})]$. Hence, we obtain the powers of the
small parameter $b_m$, that we need.
\end{Rk}
This finishes the proof of Proposition. \hfill $\square$

As a result we can write for $I_n^T$ the following
estimate:
\begin{equation}
I^T_n(\bar Y,\eta,\A_B;(s)_{n-1}) \leq C(\A_B) (8db_m\d_m)^{n-1}
\prod_{k=1}^{n-1}d_\eta(k)!\times \label{4.8}
\end{equation}
\begin{equation*}
\times\prod_{k=2}^n\Big(\sup_{t'_k}\int_{Y_k}dt_k
G_0^T(t'_k,t_k)\Big) \int_{\HC_{\hat{\b},\L}}
d\mu^{\L,X_n}_{0;(s)_{n-1}}(\p_{X_n})\prod_{k=1}^n e^{-
\d_m\p(t_k)^2 /4},\quad t_k\in Y_k.
\end{equation*}
The next step is to estimate the Gaussian integral in (4.8).

\begin{Pn}
Let
\begin{equation}
\tilde I_n= \int_{\HC_{\hat{\b},\L}}
d\mu^{\L,X_n}_{0,(s)_{n-1}}(\p_{X_n})\prod_{k=1}^n e^{-
\d_m\p(t_k)^2 /4},\quad t_k\in Y_k,\quad n\geq 2,\label{4.9}
\end{equation}
then there exists a constant $c$, which depends on parameters $d,
J, a$ such that

\begin{equation}
\tilde I_n\leq e^{c|X_n|}\quad
\Big(\frac{2\sqrt{a}}{2\sqrt{a}+\d_m}\Big)^{{d}|X_n| /4
}.\label{4.10}
\end{equation}
\end{Pn}
{\it Proof.} Write the product in (4.9) as new a Gaussian integral
w.r.t. the measure $d\g(z)=1/(2\pi)^{{dn}/{2}}\exp[- z^2 /2]$,
$z\in\R^{dn}$:

\begin{equation}
\tilde I_n= \int_{\HC_{\hat{\b},\L}}
d\mu^{\L,X_n}_{0;(s)_{n-1}}(\p_{X_n})\int_{\R^{dn}} d\g(z)
e^{i\sqrt{{\d_m}/{2}}\sum_{k=1}^nz_k\cdot\p(t_k)}.\label{4.11}
\end{equation}
Interchanging integrals one obtains

\begin{equation}
\tilde I_n =  \int_{\R^{dn}} d\g(z) e^{-{\d_m}/{4} \sum_{k,k'=1}^n
z_k z_{k'}
 G_0^{\L,X_n}(t_k,t_{k'};(s)_{n-1},0)}, \label{4.12}
\end{equation}
where  $G_0^{\L,X_n}$ is defined by (3.12). By construction
(3.9)-(3.10) the function $G_0^{\L,X_n}(t_k,t_{k'};(s)_{n-1},0)$
is a convex combination of ``diagonalized'' covariances (see [21],
or [30, Sec.18.2]):

\begin{equation}
 G_0^{\L,X_n}(t_k,t_{k'};(s)_{n-1},0)=\sum_i\l_iG_0^{(i)},\quad
\sum_i\l_i=1, \label{4.13}
\end{equation}
where coefficients $\l_i$ are monomials of the form
 $s_{k_1}\cdots s_{k_p}(1-s_{l_1})\cdots (1-s_{l_{p'}})$ and

\begin{equation}
 G_0^{(i)}=
\sum_k\1_{Z_k(i)}G_0^T\1_{Z_k(i)},\quad \bigcup_kZ_k(i)=X_n\quad
\mbox{for all}\quad i . \label{4.14}
\end{equation}
Substituting (4.13) into (4.12),  applying then the H\"older
inequality, and returning  back to the measure
$d\mu^{\L,X_n}_{G_0^{(i)}}$, defined by the covariance (4.14), one
gets:

\begin{equation*}
\tilde I_n \leq \prod_i\Big( \int_{\R^{dn}} d\g(z) e^{
-{\d_m}/{4}\sum_{k,k'=1}^n z_k z_{k'}
 G_0^{\L,X_n}(t_k,t_{k'})}\Big)^{\l_i}=
\end{equation*}

\begin{equation}
\prod_i\Big( \int_{\HC_{\hat{\b},\L}}
d\mu^{\L,X_n}_{G_0^{(i)}}(\p_{X_n})\prod_{k=1}^n e^{-\d_m
\p(t_k)^2 /4}\Big)^{\l_i}=:\prod_iK_i^{\l_i}.\label{4.15}
\end{equation}
By virtue of (4.14) we have:

\begin{equation}
K_i=\prod_k\Big( \int_{\HC_{\hat{\b},\L}}
d\mu^{\L,Z_k(i)}_{G_0^{(i)}}(\p_{X_n}) \prod_{Y\in Z_k(i)}
e^{-\d_m\p(t_Y)^2 /4}\Big)=:\prod_k J_k(Z_k(i))\label{4.16}.
\end{equation}
By definition (2.23) one gets for $\L=Z_k(i)$ and
$F(\p_{Z_k(i)})=\prod_{Y\in Z_k(i)}e^{- \d_m\p(t_Y)^2 /4 }$:

\begin{equation}
J_k(Z_k(i))=C_0(Z_k(i);J)^{-1} \int_{\Om_{\hat{\b},\L}}
d\om_{Z_k(i)} e^{-\frac{J}{4}\int_{Z_k(i)}^*dt(\om(t)-\om(t_*))^2}
\prod_{Y\in Z_k(i)} e^{- \d_m\om_Y(\tau_Y)^2 /4 }.\label{4.17}
\end{equation}
Here the  measure $d\om_{Z_k(i)}$ is defined by l.h.s. of (2.23),
with $J=0$ in definition of operator $B_\L$ (\ref{2.15}) and with
$\L_k(i):=Z_k(i)\cap\L$,\quad $S_k(i):=Z_k(i)\cap S_{\hat\b}$:
\begin{equation*}
\int_{\Om_{\hat{\b},\L}} d\om_{Z_k(i)}(...)=
 \int_{\R^{d|\L|}} dx_\L\int_{\Om_{\hat{\b},\L}}
W_{x_\L;x_\L}^{\hat{\b}}(d\om_\L)
e^{-\frac{a}{2}\int_0^{\hat{\b}}d\tau (\om_\L(\tau),\om_\L(\tau))
+\frac{d}{2} Tr B_\L}(...) \,.
\end{equation*}
We put also :

\begin{equation}
\int_{Z_k(i)}^*dt(\om(t)-\om(t_*))^2:= \sum_{\langle l,l'\rangle
\ss \L_k(i)}\int_{S_k(i)}d\tau(\om_{l}(\tau) -\om_{l'}(\tau))^{2}.\label{4.18}
\end{equation}
The normalization factor $C_0:=C_0(Z_k(i);J)$ is the integral
(4.17) calculated for $\d_m=0$. By the Cauchy-Schwarz inequality
we get from integral (4.17) that
\begin{equation}
J_k(Z_k(i))\leq \frac{C_0(Z_k(i);2J)^{1/2}}{C_0(Z_k(i);J)}
\Big(\prod_{Y\in Z_k(i)}\int_{\Om_{\hat{\b}}} d\om_Y e^{-
{\d_m}\om_Y(\tau_Y)^2 /2}\Big)^{1/2}.\label{4.19}
\end{equation}
The factors $C_0$ for $J$ and $2J$ can be calculated explicitly
using (2.24), (2.18). The integral in the product (4.19) can be
estimated (for $|Y|=1$) by

 \begin{equation*}
\int_{\Om_{\hat{\b}}} d\om_Y e^{- {\d_m}\om(t_Y)^2 /2}= C_0(Y,0)
\int_{\HC_{\hat{\b},\L}} d\mu_{G_0^\L}^Y(\p_Y) e^{- {\d_m}\p_{l_Y
/2}(\tau_Y)^2}\Big|_{J=0}=
\end{equation*}

 \begin{equation}
 C_0(Y,0)\Big(1+\frac{\d_m}{2\sqrt{a}}\cdot
\frac{e^{\hat{\b}\sqrt{a}}+1}{e^{\hat{\b}\sqrt{a}}-1}\Big)^{-d/2}
\leq e^{\mbox{const}|Y|}\Big(\frac{2\sqrt{a}}
{2\sqrt{a}+\d_m}\Big)^{d/2}.\label{4.20}
\end{equation}
This finishes the proof of the Proposition. \qquad \hfill
$\square$

Now we return to the proof of Lemma 3.1. Since

\begin{equation*}
\sum_Y\int_YdtG_0^T(t'-t)=\int_TdtG_0^T(t'-t)=
\sum_{j\in\L}\int_0^{\hat{\b}}d\tau G_{0,ij}^\L(\tau'-\tau)=
\end{equation*}
\begin{equation}
=:\sum_{j\in\L}D_{0}^\L(i-j)\leq\sum_{j\in\Z^\nu}D_{0,}(j)=:C_G
\,, \label{4.21}
\end{equation}
then taking into account dependence of $b_m$ and $\d_m$ on $m$
(see (2.10)) and definition (3.21), we obtain by summing over
\textit{trees} $\eta$ the estimate:
\begin{displaymath}
 |\KC_{\bar y}^{T}(\A_{B})|\le C'(\A_B)\big(16b\sqrt{a}C_G m^{{d}/{8}}
  \big)^{n-1}\times
\end{displaymath}
\begin{displaymath}
\times\sum_{\eta}\prod_{p=1}^{n}d_{\eta}(p)!
  \int_0^1(ds)^{n-1}f(\eta;(s)_{n-1}).
\end{displaymath}
The last step of the proof is due to the Battle--Federbush
inequality [16,17]:
\begin{displaymath}
 \sum_{\eta}\prod_{p=1}^{n}d_{\eta}(p)!\int_0^1(ds)^{n-1}
   f(\eta;(s)_{n-2})   \le 4^{n}, \qquad n = |\eta|.
\end{displaymath}
This yields (3.28) with
\begin{displaymath}
 \ve(m)= 64b\sqrt{a}C_G m^{{d}/{8}} .
\end{displaymath}
Now taking into account (3.30), we obtain the value of the
small-mass threshold $m_*$:
\begin{equation}
m_* \; = \; (64b\sqrt{a}C_Ge^c)^{-{8}/{d}}.\label{4.22}
\end{equation}
\hfill $\square$\\
{\bf Proof of Lemma 3.2.} The proof is standard. One can use
general theory of  polymer-type expansions  [42] (or,
equivalently, the \textit{Kirkwood--Salsburg} type technique [30,
Sec.18.5]), see, also, [49]. \hfill $\square$ \\
{\bf Proof of Corollary 3.1.} If we restrict the sum in the
cluster expansion (3.20)--(3.27) in such a way that, for example,
$Y_2\ss B'$,  then all terms in this expansion are proportional to
$G_0^T(t_1, t_2)$, where $t_1\in B$, $t_2\in B'$, then (3.32)
follows from the asymptotic behaviour of $G_{0,|i-j|}(\tau)$ in
the variable $|i-j|$.  Note, also, that in the case of
\textit{p.b.c.} (see e.g.[30])
\begin{equation*}
G_{0;|i-j|}(\tau\mid p)\leq G_{0;|i-j|}(\tau\mid 0),
\end{equation*}
and it has exponential decay of $G_{0;|i-j|}(\tau\mid 0)$ in
$|i-j|$, which is uniform in $\tau$. \hfill $\square$\\
{\bf Proof of Lemma 3.4.} To prove this lemma one has to check
that Propositions 4.1 and 4.2 are true for the \textit{auxiliary
potential} (3.38). It is easy to verify that (4.5)-(4.7) are true
for $X_{\pm}=X_{\pm}(x\mid y)=-b_me^{- \d_m(x\pm y)^2 /4}$ with
$\d_m/2$ instead of $\d_m$. So, the  estimate (4.4) is the same
but with $[e^{- \d_m(x(t_k)+y(t_k))^2 /8}+ e^{-
\d_m(x(t_k)-y(t_k))^2 /8}]$ instead of $e^{-\d_m\p(t_k)^2 /4}$.
Then in the proof of Proposition 4.2 we have to modify the
integral (4.20) to

\begin{equation}
\int_{\HC_{\hat{\b},\L}}  d\mu_{G_0^\L}^Y(x_Y)
e^{-\frac{\d_m}{8}(x_{l_Y}(\tau_Y)\pm y_Y(\tau_Y)) ^2}\Big|_{J=0},
\label{4.23}
\end{equation}
which gives

\begin{equation*}
C_0(Y,0)\Big(1+\frac{\d_m}{4}G_0^\L(0)\Big)^{-d/2}e^{-\frac{\d_m
y^2} {8 (1+ \d_m G_0^\L(0) /4)}}\Big|_{J=0}.
\end{equation*}
Since the last exponent is less than one, the dependence on the
configuration $y$ disappears, and this finishes the proof. \hfill $\square$\\
{\bf Proof of Lemma 3.5.} We follow the line of reasoning in the
proof of Theorem 3.2 [6]. Let $d\tilde\mu_0^\L(\p,\p'\mid 0):=
d\mu_0^\L(\p\mid 0)\otimes d\mu_0^\L(\p'\mid 0)$. Then we can
rewrite (3.46) as
\begin{equation*}
K^\L_{ll'}(\tau,\tau'\mid\eta)=Z(\L\mid\eta)^{-2}\int_{\HC_{\hat{\b},\L}}
 \int_{\HC_{\hat{\b},\L}}
\frac{\p_l(\tau)-\p'_l(\tau)}{\sqrt{2}}\cdot
\frac{\p_{l'}(\tau')-\p'_{l'}(\tau')}{\sqrt{2}}\times
\end{equation*}

\begin{equation}
\times e^{-\hat{V}(\p_\L)-\hat{V}(\p'_\L)}
e^{\frac{J}{2}\sum_{\langle
l,l'\rangle}^{\pl\L}\int_0^{\hat{\b}}d\tau
(\p_l(\tau)-\p'_l(\tau)) \eta_{l'}(\tau)}
d\tilde\mu_0^\L(\p,\p'\mid 0).\label{4.24}
\end{equation}
After change of variables:

\begin{equation*}
x_l(\tau)=\frac{\p_l(\tau)-\p'_l(\tau)}{\sqrt{2}},\quad
y_l(\tau)=\frac{\p_{l}(\tau)+\p'_{l}(\tau)}{\sqrt{2}},
\end{equation*}
by  definition of the measures it is clear that
\begin{equation}
d\tilde\mu_0^\L(\p,\p'\mid 0)= d\tilde\mu_0^\L(x,y\mid 0).\label{4.25}
\end{equation}
Then, using definition (3.44) we get:

\begin{equation}
K^\L_{ll'}(\tau,\tau'\mid\eta)=Z(\L\mid\eta)^{-2}\int_{\HC_{\hat{\b},\L}}
 \int_{\HC_{\hat{\b},\L}}
x_l(\tau)x_{l'}(\tau')
e^{-\tilde{V}(x_\L\mid y_\L)}\times
\end{equation}
\begin{equation*}
\times e^{\frac{J}{2}\sum_{\langle
l,l'\rangle}^{\pl\L}\int_0^{\hat{\b}}d\tau y_l(\tau)
\eta_{l'}(\tau)} d\tilde\mu_0^\L(x,x'\mid 0).\label{4.26}
\end{equation*}
Define the measure

\begin{equation}
\tilde\mu^\L(dx_\L\mid y_\L)=\tilde Z(y_\L)^{-1}
e^{-\tilde{V}(x_\L\mid y_\L)}d\mu_0^\L(x_\L\mid 0)\label{4.27}.
\end{equation}
Then
 \begin{equation*}
K^\L_{ll'}(\tau,\tau'\mid\eta)=Z(\L\mid\eta)^{-2}\int_{\HC_{\hat{\b},\L}}
   d\mu_0^\L(y_\L\mid 0) \tilde Z(y_\L)
e^{\frac{J}{2}\sum_{\langle l,l'\rangle}^{\pl\L}\int_0^{\hat{\b}}d\tau
y_l(\tau) \eta_{l'}(\tau)}\times
\end{equation*}
\begin{equation}
\times \langle
x_l(\tau)x_{l'}(\tau')\rangle_{\tilde\mu^\L(\cdot\mid y_\L)}.
\label{4.28}
\end{equation}
Now the proof follows from Lemma 3.4, or more explicitly from
(3.33) with $\A_{B_1}=x_l$ and   $\A_{B_2}=x_{l'}$.  \hfill
$\square$


\section{External field analyticity of the Gibbs state}

In the previous sections we have proved uniqueness of the
equilibrium state for our model (in the DLR-sense) in the set of
tempered EGM, see (3.34). In this section we consider the model
(2.1), (2.2) in the presence of external field $h$:

\begin{equation}
H^\L(h) \; = \; H_0^\L + V^\L + \sum_{j\in\L}h\cdot q_j, \quad
h\in\R^d. \label{5.1}
\end{equation}
Here $H_0^\L$ and $V^\L$ are defined  by (2.1)--(2.3). After
rescaling (2.6) one gets:
\begin{equation}
\hat{H}^\L(h) \; = \; \hat{H}_0^\L + \hat{V}^\L +\sum_{j\in\L}
\hat{h}\cdot x_j, \quad \hat{h}=m^{-1/4}h .\label{5.2}
\end{equation}
Then  the following statement is true:

\begin{Th}
For the system of quantum particles with Hamiltonian (5.1)
there is $m_*$, such that for any $0<m< m_{*}$, $T\geq 0$ and
$|h^{(\a)}|<h_0(m_*), \a=1,...,d$,  the limit
\begin{displaymath}
\lim_{\L\nearrow\Z^\nu} \r_{\L}^{\b,h}(A) \; = \; \r^{\b,h}(A), \; \;
A \in \AC_{0},
\end{displaymath}
exists. The limit $ \rho^{\b,h}(A) $ gives a unique state on the
algebra $ \AC_{0} $, which can be continuously extended to the
algebra $ \AC $. Moreover, the functions $ \rho^{\b,h}(A)\,, A \in
\AC$ are analytic in the domain $\{h^{(\a)}\in {\mathbb{C}}^d\mid
|h^{(\a)}|<h_0(m_*)\}$.
\end{Th}
{\it Proof.}  The proof is close to that of Theorem 2.1. The only
remark one has to do is that in  expansions (3.20)-(3.27) the
factor $\exp[-\hat{V}(\p_{X_n})]\exp[\hat{h} \int_{X_n}\p(t)dt]$
appears
 instead
$\exp[-\hat{V}(\p_{X_n})]$  and the operator $\D(\eta,\bar Y)$ in
(3.22) acts on the product
$\A_B(\p_B)\exp[-\hat{V}(\p_{X_n})]\exp[h_m\int_{X_n}\p(t)dt]$.
When the derivatives from $\D(\eta,\bar Y)$ act on the factor
$\exp[\hat{h}\int_{X_n}\p(t)dt]$, the convergence is guaranteed by
the small factors $h = m^{1/4}\hat{h}$ from $\{h\in
{\mathbb{C}}^d\mid |h^{(\a)}|<h_0(m_*), \a=1,...,d\}$. In
principle this implies a new value of the \textit{light-mass}
threshold $m_*^h$:

\begin{equation}
m_*^h= \min\{m_*, m_* (h C_G e^{c+1})^{-4}\},\label{5.3}
\end{equation}
where $m_*$ is defined in (4.22), $C_G$ is from the estimate of
the product of covariances in $\D(\eta,\bar Y)$, $e^c$ is from the
estimate of $F_n^T(\bar Y)$, and additional $e$ is from the
inequality (cf. [21]):

\begin{displaymath}
 \sum_{\eta}\int_0^1(ds)^{n-1}
   f(\eta;(s)_{n-2})   \le e^{n}, \qquad n = |\eta|.
\end{displaymath}
But by virtue of (5.3) there is $h_0(m_*)$ such that $m_*^h=m_*$
for $|h^{(\a)}|<h_0(m_*)$, which finishes the proof. \hfill $\square$\\
As a consequence of the above theorem we  obtain the following
result (cf. [43]):

\begin{Th}

Let $\L_n$ be the sequence $\L_1\ss\L_2\ss ...\ss\L_n...$,
 $\cup_{n=1}^\infty\L_n=\Z^\nu$, and let $h= e \|h\|$, where $e$ is
a unit vector in ${\mathbb{R}}^d$. By

\begin{equation}
\s(\b) \; := \; \lim_{h\rightarrow 0}\lim_{n\rightarrow\infty}
{\r}_\L^{{\b},h}\Big(\frac{1}{|\L_n|}\sum_{j\in\L_n}q_j\cdot e
\Big) \label{5.4}
\end{equation}
we define the order parameter for the model (5.1).  Then there
exists mass $m_*> 0$ such that for any $0 < m < m_*$ and for all
temperatures, including $\b=\infty$, the order parameter is
trivial for any tempered boundary condition defined in (3.34),
i.e.:

\begin{equation}
\s(\b) \; = \; 0.\label{5.5}
\end{equation}
\end{Th}
{\it Proof.} The proof follows directly from the Theorem 5.1 and
cluster expansion for the state

\begin{equation}
\hat{\r}_\L^{\hat{\b},\hat{h}}\Big(\frac{1}{|\L_n|}\sum_{j\in\L_n}x_j\cdot
e \Big). \label{5.6}
\end{equation}
Indeed, if derivatives from\quad $\D(\eta,\bar Y)$ act on the
factor
\begin{equation*}
 \exp\left\{\int_{X_n}\hat{h}\cdot \p(t)dt\right\}\,,
\end{equation*}
then after thermodynamic limit the corresponding terms are
proportional to $h = m^{-1/4}\hat{h}$. If derivatives act on the
factor $\p_l\exp[-\hat{V}(\p_{X_n})]$, then in the limit we obtain
a Gaussian integration with zero mean and with odd numbers of
fields. So, taking into account that ${\r}^{{\b},{h}}$ is analytic
in ${h}$ (Theorem 5.1), in the limit $h=0$ we get zero value (5.5)
for the order parameter. \hfill $\square$


\section{Convergence of cluster expansions and Gibbs state uniqueness for high temperatures}

The problem of the existence and uniqueness of the quantum state
(EGM) at high temperatures for the model (2.1)-(2.3) can be solved
in the framework of a general approach, developed e.g. in [13,14].
But there exist an independent question of construction of high
temperature expansions for quantum state (2.5) and of the proof of
its convergence in thermodynamic limit. This question is related
to a specificity of the one-site potential (2.3) behaviour at
infinity. For the quantum systems with unbounded spins the
corresponding expansions were considered in [35,36,45,46,53]. All
these results were obtained for the models with the one-site
polynomial anharmonic potentials of the form (1.3). Then the
convergence is insured by the small parameter $\b^{(s-1)/{2s}}$.
Since in (2.1)-(2.3) we have $s=1$, there is no such parameter in
our case. We recall that $2s$ is degree of one-site anharmonic
polynomial interaction in (1.3).

In the present section we show that approach we developed in
Sections 3--5 (the choice of reference measure and the type of
cluster expansion) can be adapted to obtain convergent cluster
expansions in powers of $\b > 0$, and to prove analyticity of the
state in domain $|\b - \b_*/2| < \b_*/2$ of the complex values of
$\b$ around the point $\b_*/2> 0$.

The construction of cluster expansions is absolutely the same as
above, but instead of partition of the interval  $[0,\b]=S_\b$
into unit intervals we consider it as  \textit{one unit} with
variable $\tau$ running in $S_\b$. In another words, instead of
"rods" $\D_{j,\hat{\tau}}$ (see (3.5)) we consider for every site
$j\in\L$ we consider "rods" $\D_{j,\b}:=(j,S_\b)$. Then the
measure $d\mu_{0;(s)_n}^\L$ is constructed in the same way as in
Section 3 with help of the covariance $G_{0;j,k}^\L(\tau,
\tau';(s)_n)$, see (3.12), where $X_n = \emptyset$ and $T =
\Lambda$ . Then following the line of reasoning of Section 3 we
obtain:

\begin{equation}
\langle\A_B( \cdot )\rangle_{\mu^{\L}}=
\sum_{n=1}^{n_\L}\sum_{\bar Y\ss \L} \KC^\L_{\bar Y}(\A_B)
F^\L_n(\bar Y),\label{6.1}
\end{equation}
with the same definitions as in (3.24)-(3.30), and with only
difference that

\begin{equation}
\D_{p,p'}^{(\a)}=\int_0^{{\b}}d\tau\int_{0}^{{\b}}d\tau^{'}
G_{0;Y_p,Y_{p'}}^\L(\tau,\tau^{'})
\frac{\d^2}{\d\p^{(\a)}_{Y_p}(\tau)\d\p^{(\a)}_{Y_{p'}}(\tau^{'})},\label{6.2}
\end{equation}
and $Y_p$, $Y_{p'}$ are the sites of the lattice $\Z^\nu\cap\L$.
The main result then is the following

\begin{Th}
For the quantum system with Hamiltonian (2.1)--(2.3) there exists
a sufficiently  small (independent of mass m) value of inverse
temperature $\b_{*}$ such that for all $ 0 < \b <  \b_{*} $ the
weak limit of the measures
\begin{displaymath}
\lim_{\L \nearrow \Z^\nu} \mu^\L \; = \; \mu,
\end{displaymath}
exists and $\mu$ is a unique translation-invariant Gibbs measure
on ($\Om_{{\b},\Z^\nu},\S_{{\b},\Z^\nu}$) in the set of tempered
measures. Here $\mu^\L$ as in Theorem 2.2.
\end{Th}
{\it Proof.} It is very close to the proof of Theorem 2.2. The
only difference is that in the estimate (4.20) one should take

\begin{equation}
\Big(1+\frac{\d_m}{2\sqrt{a}}\cdot
\frac{e^{\hat{\b}\sqrt{a}}+1}{e^{\hat{\b}\sqrt{a}}-1}\Big)^{-d/2}
\leq \Big(\frac{2\hat{\b}\sqrt{a}}
{2\hat{\b}\sqrt{a}+\d_m}\Big)^{d/2}\leq (\frac{2\b\sqrt{a}}{\d})^{d/2}.\label{}
\end{equation}
This gives for the small parameter of cluster expansion the value

\begin{equation}
\ve \; = \; 64b\sqrt{a}C_G\b^{d/2}.
\end{equation}
Therefore,
\begin{equation}
\b_*\; = \; (64b\sqrt{a}C_G e^c)^{-{2}/{d}},
\end{equation}

\hfill $\square$


\section{Conclusions}

In the present paper we construct the Gibbs states for quantum
crystal with nonpolynomial anharmonic potential (1.5) and with
nearest neighbour harmonic interaction in the \textit{light-mass
regime}. These states are constructed on the algebra of
quasi-local observables described in Section 2. Our analysis is
based on a reformulation of the quantum Gibbs states problem into
the study of Euclidean Gibbs Measures (EGM) [2,7,31,33].

The main technical tool we use to get the main result is the
\textit{cluster expansions} for EGM. They allow us (Theorem 2.1)
to prove the existence and \textit{uniqueness} of
translation-invariant EGM in the \textit{light-mass} domain, for
all temperatures $T$, including $T=0$ ($\b=\infty$).

To prove the existence of a unique translation-invariant measure
we start with  \textit{periodic } boundary conditions (Section 2).
Then we prove convergence of the corresponding cluster expansions
and the cluster property of the EGM in the \textit{light-mass}
domain for all temperatures $T$. This gives the
\textit{uniqueness} in the class of translation-invariant measures
for small mass $m < m_*$ and all $T\geq 0$. Next we establish that
convergence of the EGM cluster expansions and their cluster
properties hold also for \textit{zero } boundary conditions. This
again implies uniqueness in this case. Finally, following the
arguments of [6,11] we prove the uniqueness of EGM on the set of
\textit{tempered} configurations (Section 3).


\vskip5mm

{\bf Acknowledgments}. We would like to thank S. ~Albeverio, Yu.
~Kondratiev, Yu. ~Kozitsky, and M. ~R\"ockner for sending us their
article [6] prior to publication and for fruitful discussions.
One of the author (A.L.R.) would like to thank the Centre de
Physique Th\'{e}oretique, CNRS-Luminy for the warm hospitality
during his stay in October--January of 2002-2003.

\section*{References}

\begin{itemize}

\item[1.]  V.L.~Aksenov, N.M.~Plakida, and S.~Stamenkovi\'c, \textit{%
Neutron Scattering by Ferroelectrics}, World Scientific,
Singapore 1990.

\item[2.]  S.~Albeverio and R.~H\o egh-Krohn, Homogeneous random fields
and quantum statistical mechanics, {\it J.\ Funct.\ Anal.},
\textbf{19}, 242--272 (1975).

\item[3.]  S.~Albeverio, Yu.~Kondratiev, and Yu.~Kozitsky, Suppression of Critical
Fluctuations by Strong Quantum Effects in Quantum Lattice Systems,
 {\it Commun. Math. Phys.}, \textbf{194}, 493--512 (1998).

\item[4.]  S.~Albeverio, Yu.~Kondratiev, Yu.~Kozitsky, and M.~R\"ockner,
Uniqueness for Gibbs mearsures of quantum lattices in small mass
regime, {\it Ann. Inst. H. Poincare}, \textbf{37}, 43--69 (2001).

\item[5.]  S.~Albeverio, Yu.~Kondratiev, Yu.~Kozitsky, and M.~R\"ockner,
 Gibbs states of quantum crystal:~uniqueness by small particle mass.
{\it C.R. Acad. Sci. Paris, Ser. I}, \textbf{335}, 693--698
(2002).

\item[6.]  S.~Albeverio, Yu.~Kondratiev, Yu.~Kozitsky, and M.~R\"ockner,
Small mass implies uniqueness of Gibbs states of a quantum
crystal, {\it Commun. Math. Phys.}, \textbf{241}, 69--90 (2003).

\item[7.]  S.~Albeverio, Yu.~Kondratiev, Yu.~Kozitsky, and M. R\"ockner,
Euclidean Gibbs states of Quantum Lattice Systems, {\it Rev. Math.
Phys.}, {\bf 14}, 1335--1401 (2002).

\item[8.]  S.~Albeverio, Yu.G.~Kondratiev, R.A.~Minlos, and
A.L.~Rebenko, Small mass behaviour of quantum Gibbs states for
lattice models with unbounded spins, {\it J. Stat. Phys.}, {\bf
92}, 1153--1172,
 1998.

\item[9.]  S.~Albeverio, Yu.G.~Kondratiev, R.A.~Minlos, and
G.~V.~Shchepan'uk, Uniqueness Problem for Quantum Lattice Systems
with  Compact Spins, {\it Lett. Math. Phys.},  {\bf 52}, 185--195,
2000.

\item[10.]  S.~Albeverio, Yu.~Kondratiev, T. Pasurek, and M. R\"ockner,
Gibbs states on loop lattice: existence and a priori estimates,
{\it C. R. Acad. Sci. Paris,} {\bf 333}, Serie I 1005--1009 (2001)

\item[11.]  S.~Albeverio, Yu.~Kondratiev, T. Pasurek, and M. R\"ockner,
A priori estimates  and existence for Euclidean Gibbs measures,
{\it  Preprint, BiBoS, Nr 02-06-089, Bielefeld (2002)}

\item[12.]  S.~Albeverio, A.Yu.~Kondratiev, and A.L.~Rebenko, Peierls
argument and long-range order behaviour of quantum lattice systems
with unbounded spins, {\it J. Stat. Phys.}, {\bf 92}, 1137--1152,
1998.

\item[13.]  S.~Albeverio, Yu.G.~Kondratiev, M.~R\"ockner, and
T.V.~Tsikalenko, Uniqueness of Gibbs States for Quantum Lattice
Systems, {\it Probab.\ Theory Relat. Fields}, \textbf{108},
193--218 (1997).

\item[14.] S.~Albeverio, Yu.G.~Kondratiev, M.~R\"ockner, and
T.V.~Tsikalenko, Dobrushin's Uniqueness for Quantum Lattice
Systems with Nonlocal Interactions, {\it Commun. Math. Phys.},
\textbf{189}, 621--630 (1997).

\item[15.]  V.S.~Barbulyak and Yu.G.~Kondratiev, Functional Integrals and
Quantum Lattice Systems: III~Phase transitions. {\it Reports Nat.\
Acad.\ Sci.\ of Ukraine,} No~10, 19--21 (1991).

\item[16.] G.~A.~Battle III, A new combinatoric estimate for cluster
expansions, {\it Commun.\ Math.\ Phys.}, {\bf 94}, 133--139
(1984).

\item[17.] G.~A.~Battle III and P.~Federbush, A note on cluster expansions,
tree graph identities, extra $1/N!$ factors!!!, {\it Lett.\ Math.\
Phys.}, {\bf 8}, 55--57 (1984).

\item[18.]  O.~Bratteli and D.W.~Robinson, \textit{Operator algebras and
Quantum Statistical Mechanics}, Vol.~2 (second edition),
Springer-Verlag, Berlin 1996.

\item[19.] A.~D.~Bruce and R.~A.~Cowley, {\it Structural phase transitions,}
Taylor and Francis Ltd, London 1981.

\item[20.] D.~Brydges, A rigorous approach to Debye screening in dilute
classical Coulomb systems, {\it Commun.\ Math.\ Phys.}, {\bf 58},
313--350 (1978).

\item[21.] D.~Brydges and P.~A.~Federbush, A new form of the Mayer expansion
in classical statistical mechanics, {\it J.~ Math.\ Phys.}, {\bf
19}, 2064--2067 (1978).

\item[22.] D.~Brydges and P.~A.~Federbush, Debay Screneeng,
{\it Commun. \ Math.\ Phys.} {\bf 73}, 197--246 (1980).

\item[23.] V.~Cachia and V.~A.~Zagrebnov, Trotter product formula
for nonself-adjoint Gibbs semigroups, {\it J.~ Lond.\ Math.\
Soc.}, {\bf 64}, 436--444 (2001).

\item[24.] Ph. Choquard, {\it The Anharmonic Crystal}, W. A. Benjamin,
N.Y. 1967.

\item[25.] R.L. Dobrushin, Prescribing a System of Random Variables
by Conditional Distributions, {\it Theory Prob. Appl.}, {\bf 15},
458--486 (1970).

\item[26.]  W.~Dressler, L.~Landau, and J.F.~Perez, Estimates of critical
length and critical temperatures for classical and quantum lattice
systems, {\it J.\ Stat.\ Phys.}, \textbf{20}, 123--162 (1979).

\item[27.] W.G. Faris and R.A. Minlos, A Quantum Crystal with
Multidimensional Anharmonic Oscillators, {\ it J. Stat. Phys.},
{\bf 94}, 365--387 (1999).

\item[28.] J. Ginibre, General Formulation of Griffiths' Inequalities,
{\it Commun. Math. Phys.}, {\bf 16}, 310--328 (1970).

\item[29.] J.~Glimm, A.~Jaffe,~ and T. Spencer, The particle structure of
the weakly coupled $P(\p)_2$ model and other applications of high
temperature expansions. Part II. The cluster expansion. In: {\ it
Constructive quantum field theory} (eds. G. Velo, A. Wightman).
{\it Lecture Notes in Physics} {\bf 25}, Springer-Verlag, Berlin,
Heidelberg 1973.

\item[30.] J.~Glimm and A.~Jaffe, {\it  Quantum physics. A functional
integral point of view}, Springer-Verlag, Berlin 1987.

\item[31.]  S.A.~Globa and Yu.G.~Kondratiev, The construction of Gibbs
states of quantum lattice systems, {\it Selecta Math.\ Sovietica},
\textbf{9}, 297--307 (1990).

\item[32.]  R.~H\o egh-Krohn, Relativistic Quantum Statistical
Mechanics in Two-Dimensional Space-Time, {\it Commun.\ Math.\
Phys.}, \textbf{38}, 195--224 (1974).

\item[33.]  A.~Klein and L.~Landau, Stochastic processes associated with
KMS states, {\it J.\ Funct.\ Anal.}, \textbf{42}, 368--428 (1981).

\item[34.]  Yu.G.~Kondratiev, Phase transitions in quantum models of
ferroelectrics, in: {\it  Stochastic proceses, physics and geometry},
World Scientific, Singapore, N.Jersey, 465--475 (1994).

\item[35.]  A.~Yu.~Kondratiev and A.~L.~Rebenko, Some remarks about
 cluster expansions for ubounded continuous spin systems in
quantum statistical mechanics, {\it Methods Funct.\ Anal.\ and
Topology}, {\bf 2},  83--99 (1996).

\item[36.] A.~Yu.~Kondratiev and A.~L.~Rebenko, Cluster expansions of
Brydges--Federbush type for quantum lattice systems, {\it Methods
Funct.\ Anal.\ and Topology}, {\bf 2},  59--68 (1996).

\item[37.] Yu. Kozitsky, Quantum Effects in a Lattice Model of
Anharmonic Vector Oscillators, {\it Lett. Math. Phys.}, {\bf 51},
71--81 (2000).

\item[38.] Yu. Kozitsky, Scalar Domination and Normal Fluctuation in
N-Vector Quantum Anharmonic Crystals, {\it Lett. Math. Phys.},
{\bf 53}, 289--303 (2000).

\item[39.] Yu. Kozitsky, Quantum Effects in a Lattice Model of
Vector Anharmonic Oscillators, in:  {\it Stochastic Processes,
Physics and Geometry: New Interplays, II (Leipzig, 1999),}
403--411, CMS Conf. Proc., 29, {\it Amer. Math. Soc., Providence,
RI, (2000).}

\item[40.]  H.~Kunz and B.~Payandeh, Existence of phase transition for a
class of ferroelectric models near the displacive limit, {\it
Phys.\ Rev. B}, \textbf{18}, 2276-2280 (1978).

\item[41.] J.L. Lebowitz and E. Presutti, Statistical mechanics of
systems of unbounded spins, {\it Commun. Math. Phys.}, {\bf 50},
195--218 (1976).

\item[42.] V.~A.~Malyshev and R.~A.~Minlos, {\it  Gibbs random fields. The
cluster expansion method}, Kluwer, Dordrecht (1991).

\item[43.] R.~A.~Minlos, E. A. Pechersky, and V. A. Zagrebnov,
Analyticity of the Gibbs State for a Quantum Anharmonic Crystal:
No Order Parameter, {\it Annal. Henri Poincare},  {\bf 3},
921--938 (2002).

\item[44.] R.~A.~Minlos,  A.~Verbeure, and V.A.~Zagrebnov, A Quantum Crystal
Model in the Light-Mass Limit: Gibbs States, {\it Rev.\ Math.\
Phys.}, \textbf{12}, 981--1032 (2000).

\item[45.] Y. M. Park, The Cluster Expansion for Classical and Quantum
Lattice Systems, {\it J. Stat. Phys.}, {\bf 27}, 553-576 (1982).

\item[46.]  Y.M.~Park and H.J.~Yoo, A Characterization
of Gibbs States for Lattice Boson  Systems, {\it J.\ Stat.\
Phys.}, \textbf{75}, 215-239 (1994).

\item[47.]  L.A.~Pastur and V.A.~Khorushenko, Phase transition in quantum
models of rotators and ferroelectrics, {\it Theor.\ and Math.\
Phys.}, \textbf{73}, 111--124 (1987) (English translation).

\item[48.]  N.M.~Plakida and N.S.~Tonchev, Exactly soluble $d$-dimensional
model of a structural phase transition,{\it Theor.\ \&\ Math.\
Phys.}, \textbf{63}, 504--510 (1985).

\item[49.] A.~L.~Rebenko, Mathematical foundations of equilibrium
classical statistical mechanics of charged particles, {\it Russ.\
Math.\ Surveys}, {\bf 43}, 55--97 (1988).

\item[50.] G. Roepstorff, \textit{Path Integral Approach to Quantum Physics},
Springer-Verlag, Berlin 1991.

\item[51.] T.~Schneider, H.~Beck, and E.~Stoll, Quantum effects in an
$n$--component  vector model for structural phase transitions,
{\it Phys.\ Rev.}, {\bf B 13}, 1123--1130 (1976).

\item[52.] B. Simon, \textit{Functional Integration and Quantum
Physics}, Academic Press, New York 1979.

\item[53.] W.I. Skrypnik, LRO in lattice systems of linear oscillators
with strong bilinear pair nearest-neighbour interaction, {\it J.
Phys. A: Math. Gen.}, {\bf 32}, 7039--7048 (1999).

\item[54.] S.~Stamenkovi\'c, N.S.~Tonchev, and V.A.~Zagrebnov, Exactly
soluble model for structural phase transition with a Gaussian type
anharmonicity, {\it  Physica A},  \textbf{145}, 262--272 (1987).

\item[55.] J.~E.~Tibballs, R.~J.~Nelmes, and G.~J.~McIntyre, The crystal
structure of tetragonal $KH_{2}PO_{4}$ and $KD_{2}PO_{4}$ as a
function of temperature and pressure, {\it J.~Phys.\ C: Solid
State Phys.}, {\bf 15}, 37--58 (1982).

\item[56.] V.G.~Vaks, \textit{Introduction to the microscopic theory of
ferroelectrics}, Nauka, Moscow 1973 (in Russian).

\item[57.] A.~Verbeure and V.A.~Zagrebnov, Phase transitions and algebra of
fluctuation operators in an exactly soluble model of a quantum
anharmonic crystal, {\it J.\ Stat.\ Phys.}, \textbf{69}, 329--359
(1992).

\item[58.]  A.~Verbeure and V.A.~Zagrebnov, No-go theorem for quantum
structural phase transitions, {\it J.\ Phys.~A: Math.\ Gen.},
\textbf{28}, 5415--5421 (1995).

\end{itemize}

\end{document}